\documentclass[11pt,3p,review,authoryear]{elsarticle}

\usepackage{microtype}
\usepackage{graphicx}
\usepackage{subfig}
\usepackage{booktabs}
\usepackage{appendix}
\usepackage{multicol}	
\usepackage{multirow} 
\usepackage{hyperref}
\usepackage{amsmath}
\usepackage{amssymb}
\usepackage{amsthm}
\usepackage{nccmath}
\usepackage{mathtools}
\usepackage{subfig}
\usepackage{adjustbox}

\journal{International Journal of Forecasting}
\biboptions{longnamesfirst}

\begin{document}

\begin{frontmatter}

\title{Analyzing the retraining frequency of global forecasting models: towards more stable forecasting systems.}

\author[a]{Marco Zanotti\corref{cor}} 
\ead{m.zanotti22@campus.unimib.it}
\address[a]{Department of Economics, Management and Statistics, University of Milano-Bicocca, Milan, Italy}
\cortext[cor]{Corresponding author}

\begin{abstract}
    Forecast stability, that is, the consistency of predictions over time, is 
    essential in business settings where sudden shifts in forecasts can disrupt 
    planning and erode trust in predictive systems. Despite its importance, 
    stability is often overlooked in favor of accuracy. 
    In this study, we evaluate the stability of point and probabilistic 
    forecasts across several retraining scenarios using three large forecasting
    datasets and ten different global forecasting models. 
    To analyze stability in the probabilistic setting, we propose a new 
    model-agnostic, distribution-free, and scale-free metric that measures
    probabilistic stability: the Scaled Multi-Quantile Change (SMQC).
    The results show that less frequent retraining not only preserves but 
    often improves forecast stability, challenging the need for frequent 
    retraining. Moreover, the study shows that accuracy and stability are 
    not necessarily conflicting objectives when adopting a global modeling
    approach.
    The study promotes a shift toward stability-aware forecasting practices, 
    proposing a new metric to evaluate forecast stability effectively in 
    probabilistic settings, and offering practical guidelines for building more 
    stable and sustainable forecasting systems.
\end{abstract}


\end{frontmatter}


\section{Introduction} \label{sec:intro}

In recent years, global forecasting models, those trained across multiple time series 
simultaneously, have emerged as a powerful alternative to traditional local approaches, 
particularly in large-scale applications such as retail demand and energy forecasting. 
While their accuracy and efficiency have been widely studied, less attention has been 
paid to the stability of their forecasts over time.
In many operational contexts, forecasts are produced regularly and serve as the foundation 
for critical business decisions, from inventory planning to resource allocation. In such 
settings, forecast stability becomes a key requirement; not only should forecasts be 
accurate, but they should also remain reasonably stable over time.
Indeed, forecast stability refers to the consistency of predictions produced by a 
forecasting model as new data becomes available. 
As defined by \citet{stab1}, forecast stability can be categorized into two primary forms: 
vertical stability, which concerns the consistency of forecasts for the same target date 
across different forecast origins, and horizontal stability, which addresses the smoothness 
of forecasts across the forecast horizon from a single origin. Both types play a distinct 
role: vertical stability helps avoid costly forecast revisions that can disrupt planning 
cycles, while horizontal stability prevents erratic behavior across time steps that could 
lead to inefficient operations and amplification of demand fluctuations, such as the 
bullwhip effect in supply chains \citep{bullwip}.
In particular, vertical stability may also be seen as temporal robustness, in the sense 
that a forecasting model is stable to updates (or retraining) as new observations are 
available, a property that is critical in practice for avoiding frequent changes in 
business decisions. 
Forecast instability, indeed, can have serious consequences. Unstable forecasts may lead 
to frequent and expensive adjustments in supply chain plans, diminish trust in the 
forecasting system, and complicating the decision-making processes, leading to suboptimal 
business outcomes.

Nevertheless, despite the importance of forecast stability, it is still common practice
to evaluate forecasting models only on their accuracy, even in frequent retraining settings 
where stability is more likely to be compromised. This can also be due to an inherent 
perception that there exists a trade-off between accuracy and stability. As new information 
becomes available, updated forecasts may naturally differ from earlier versions, ideally 
improving in accuracy at the expense of stability. Conversely, overly stable forecasts 
might sacrifice accuracy by ignoring new, valuable information. 
Ensembling techniques, which integrate multiple forecasts from different models, have been 
proposed as a mechanism to concurrently reduce bias and variance \citep{forecombreview}. 
By aggregating diverse predictions, ensembling methods can alleviate the impact of 
individual model volatility and potentially achieve a more favorable equilibrium between
accuracy and stability, but at the expense of a severe increase in computational costs.
Even if the trade-off between forecasting accuracy and stability presents a significant 
challenge for organizations seeking to balance predictive performance with operational 
stability, recent research suggests that high levels of stability and accuracy can be 
achieved at the same time (\citep{stab1}, \citep{stab2}). 
Moreover, \citet{zanotti1} demonstrated that the retraining frequency does not harm the 
forecast accuracy of global models, suggesting that, when the data-generating process 
remains relatively stable, global models can preserve predictive performance even when 
updated less frequently. This finding raises a natural question regarding forecast 
stability. Since retraining inherently produces new model artifacts, the frequency with 
which models are retrained may influence the temporal consistency of forecasts. Whether 
less frequent retraining systematically improves, worsens, or leaves unchanged the 
stability of global forecasting models is therefore an important question that needs
to be addressed.


\subsection{Research Question}

We aim to address the question \textit{"How does the retraining frequency of global 
forecasting models affect the stability of their forecasts?"}. Purposely, we 
study whether the global modeling approach produces stable forecasts, and we try to 
understand the effects of retraining on the forecast stability, that is, whether 
avoiding re-estimation for every new observation damages the stability of global 
models. To address this question, we rely on three of the most recent and 
comprehensive forecasting datasets: the M4, the M5, and the VN1 competition data.

To generally understand the stability of the forecasting models, we consider 
ten distinct global forecasting methods (five from the traditional machine learning 
domain and five based on commonly used deep neural network architectures). We also 
compared the results with two well-known univariate local forecasting models.
Moreover, we examine stability across a range of retraining scenarios, continuous 
retraining to no retraining, by exploring periodic strategies that broadly 
encompass the most practical and effective approaches. 


\subsection{Contributions}

Our contribution is fourfold:

\begin{itemize}

    \item We provide the first comprehensive study of the forecast stability of global models, 
    using 10 distinct methods, a diverse collection of real-world datasets, and evaluating 
    both point and probabilistic predictions. 
    
    \item We analyze the relationship between the retraining period and forecast stability, 
    comparing different scenarios to quantify the impact of frequent retraining in 
    terms of the stability of forecasting.
    
    \item We suggest new metrics to evaluate the stability of probabilistic forecasts.

    \item We present practical guidelines for organizations and practitioners on when and 
    how often to retrain global forecasting models to obtain stable forecasts.    
    
\end{itemize}

By tackling these aspects, this paper contributes to both the forecasting and machine 
learning communities by providing insights into the stability of global forecasting models.


\subsection{Overview}

The rest of this paper is organized as follows. 
After a brief review of related works (Section \ref{sec:literature}), in Section 
\ref{sec:exp_des} we describe the design of the experiment used in our study.
The datasets and their characteristics are presented in \ref{s_sec:data}, and
the methods adopted for global forecasting are discussed in \ref{s_sec:forec_mod}. 
The concepts related to model update and retrain scenario are explained in 
\ref{s_sec:eval_setup}, together with the evaluation strategy adopted, while the 
metrics used to assess the stability of models are shown in \ref{s_sec:perf_met}. 
In Section \ref{sec:results} we discuss the empirical findings of our study, 
including forecast accuracy and stability on the different scenarios.
Finally, Section \ref{sec:conc} contains our summary and conclusions.


\section{Related works} \label{sec:literature}


\subsection{Global modeling}

The cross-learning approach has seen substantial development in recent years. Today, most 
time series forecasting studies include at least a benchmark comparison involving global 
models (GMs), underscoring their growing importance in the field. 

We define \(\mathcal{Y}\) as the set of all 
available time series in a dataset, where each \( Y_i \in \mathcal{Y}\)
represents an individual time series. Let \(\mathcal{F}\) denote the set of 
possible predictive functions, with each \( F \in \mathcal{F}\) corresponding to a 
specific forecasting model. 
Without loss of generality, we assume that all necessary information for prediction 
is contained within \(\mathcal{Y}\).
Under the local modeling paradigm (LMs), forecasts for a horizon \(h\) are generated by 
training a separate model for each individual time series. This implies that each 
series \(Y_i\) is associated with its own model, characterized by its own set of 
parameter values.

\begin{equation}
   Y_i^h = F(Y_i, \theta_i) \mbox{.}
   \label{eq:localmodel}
\end{equation}

In contrast, under the global modeling framework, forecasts for each individual 
time series are generated using a single model trained on the entire dataset. 
This approach leverages cross-series information, allowing the model to learn 
shared patterns and structures across all time series in \(\mathcal{Y}\).

\begin{equation}
   Y_i^h = F(\mathcal{Y}, \Theta) \mbox{.}
   \label{eq:globalmodel}
\end{equation}

It is important to note that in the cross-learning methodology, the model 
parameters \(\Theta\) are not specific to individual time series but are shared 
across all series in the dataset. This parameter sharing is a key characteristic 
of global models, enabling them to generalize patterns across series and 
potentially reducing the forecast instability.

\citet{investcrosslearn} 
demonstrated the high accuracy of GMs on the M4 competition dataset, while \citet{globsim} 
explored the conditions under which global models are competitive. Additionally, 
\citet{princlocglob} and \citep{princlocglob2} provided theoretical support showing that 
GMs can match or surpass the accuracy of LMs, with lower complexity and without 
assuming data similarity.
Global models have proven to be the most accurate method in several forecasting domains, 
including retail demand (\citet{globsales1}, \citet{globsales2}, \citet{globsales3}), 
electricity demand \citep{globelec}, water demand \citep{globwater}, gas consumption 
\citep{globgas}, and crop production \citep{globcrop}. Their effectiveness was particularly 
evident during the M5 competition \citep{m5acc}, where tree-based models employing 
cross-learning ranked among the top-performing solutions \citep{fortrees}.
To further enhance GM performance, several strategies such as clustering (\citet{globclus1}, 
\citet{globclus2}) and data augmentation \citep{globaug} have been explored. Moreover, new 
machine learning \citep{globtree} and deep learning \citep{nbeats} architectures have been 
specifically designed to support cross-learning. Recently, research has begun focusing on 
improving GMs' ability to capture local patterns \citep{globlocalize} and enhance their 
explainability \citep{globexplain}.


\subsection{Forecast stability}

From a forecasting stability perspective, the literature is impoverished. \citet{stab1}
introduced the first classification of the different types of forecasting stability, 
that is horizontal or vertical stability, proposing a new model-agnostic framework based 
on linear combinations of predictions to obtain more stable forecasts. \citet{stab2}, 
instead, extended an existing deep learning architecture (NBEATS) to optimize forecasts
from both a traditional forecast accuracy perspective, as well as a forecast stability 
perspective, directly including an instability component into the loss function of the 
model. There is also active research on forecast stability within judgmental forecasting 
ground \citep{stab3}.
However, almost all the literature related to forecast stability is focused on point 
prediction stability only. This is not due to the absence of probabilistic distance 
measures, nor to the lack of probabilistic forecasting techniques \citep{m5unc}. 
Rather, probabilistic stability has received limited attention because existing 
evaluation metrics are typically designed to compare forecasts to realizations or to 
assess calibration and sharpness, rather than to quantify temporal consistency across 
successive forecast origins.
Nevertheless, in many forecasting applications, such as supply chain management, it is 
crucial to generate and evaluate predictions probabilistically, whether through 
prediction intervals, quantiles, or full predictive distributions \citep{retailfor}. 
The seminal work of \citet{stab4} is the only one addressing the problem of probabilistic 
stability evaluation, but it only applies to a specific forecasting model (NBEATS), which
has been modified to output a full predictive Gaussian distribution.
Therefore, the stability evaluation of probabilistic forecasts is an open question, 
especially in a model-agnotic way.
Luckily, among the methods developed for uncertainty quantification, \citet{vovk2005} 
introduced Conformal Inference, which is a model-agnostic framework that offers valid 
uncertainty estimates and can also be applied in time series forecasting settings
\citep{conformalts}.


\subsection{Model retraining}

In the context of model retraining and updating strategies, the most comprehensive work in time 
series forecasting is by \citet{zanotti1}, who showed the positive effects of reducing
retraining of global models on both accuracy and forecasting costs. 
The seminal work of \citet{tsupdate1} extensively investigated the impact of various retraining 
scenarios and parameter update methods on model performance, focusing on the exponential
smoothing family of models, following the traditional local modeling approach. 
\citep{tsupdate2} touched on retraining within the retail demand domain, but the study was
restricted to a small set of models, limited retraining strategies, and a proprietary daily
dataset. While these findings are encouraging, there has been little direct investigation 
into whether global models specifically can retain stability with less frequent updates.

Building on these prior studies, our work directly examines the stability of global models. 
Moreover, by systematically evaluating a wide range of retraining strategies and their effects 
on the forecasting stability of various global models, we seek to offer both theoretical 
insights and practical guidance for promoting more stable forecasting practices.


\section{Experimental design} \label{sec:exp_des} 

This section presents the empirical analysis carried out to investigate the stability of 
global models and to determine whether less frequent retraining scenarios can yield 
stability outcomes comparable to those of the baseline scenario, which involves the most 
frequent retraining. We begin by describing the datasets used in our experiments, followed 
by an overview of the machine learning and deep learning models employed. 
Finally, we detail the instability measures, the various retraining scenarios considered, 
and the evaluation strategy applied to assess forecast performance.


\subsection{Datasets} \label{s_sec:data}

For our experiments, we employed three forecasting datasets: the M4, the M5, 
and the VN1 competition datasets. 
The M4 competition is part of the well-known M-competitions series organized 
in 2018 by Spyros Makridakis and colleagues \cite{m4comp}. It includes 
real-world time series drawn from diverse domains such as finance, economics, 
demographics, industry, and other indicators, and spans multiple frequencies. 
In our study, we focused on the 4,227 daily time series characterized by 
substantial variability in length, seasonality, and noise structure. The 
heterogeneous nature of the series makes it particularly well-suited for 
studying general retraining behaviors that are not domain-specific.
The M5 competition, also part of the M-competitions series, was organized to
benchmark forecasting methods in the context of retail demand \citep{m5comp}. 
The M5 dataset \citep{m5data} includes 30,490 daily time series representing 
unit sales of Walmart products across three categories (Food, Hobbies, and 
Household) sold in ten stores located in California, Texas, and Wisconsin. 
The data span from 2011 to 2016 and are characterized by high intermittency 
and a hierarchical structure, enabling forecasts at multiple aggregation 
levels (e.g., SKU, product category, store, and state). The dataset also 
includes exogenous variables such as prices, promotions, and special events 
(e.g., holidays), which can influence demand.
The VN1 Forecasting - Accuracy Challenge, organized in October 2024 by Flieber, 
Syrup Tech, and SupChains, represents the first edition of a new competition 
series \citep{vn1data}. The dataset comprises weekly sales data for 15,053 
products sold by U.S.-based e-vendors between 2020 and 2024. Unlike the M5, 
which involves a single retailer (Walmart) and a limited number of physical 
stores, the VN1 dataset captures sales from 328 warehouses operated by 46 
different retailers. As far as we are aware, our study is among the first to 
evaluate forecasting models on this dataset. Together, the M5 and the VN1 
datasets offer the most recent and comprehensive time series collections 
related to retail demand, enabling good generalizability of our findings
also to such domain.


\subsection{Forecasting models} \label{s_sec:forec_mod}

In our study, we focused on analyzing the stability of global methods, 
as our primary goal is to assess whether this modeling approach, unlike the 
traditional local one, can maintain performance with less frequent retraining.
Nevertheless, we also compared the results with two popular local models,
that is the Exponential Smoothing (ETS) \citep{ets} and the Autoregressive
Integrated Moving Average (ARIMA) \citep{autoarima} methods in their automatic 
versions
\footnote{
A comprehensive analysis of retraining effects on the stability of local 
models was out of the scope of this study.
}.
To conduct a comprehensive evaluation of the stability of global models, we 
incorporated both traditional machine learning algorithms and state-of-the-art 
deep learning architectures. The selected models are well-established in the time 
series forecasting literature and represent a range of methodological paradigms, 
allowing for a broad and informative comparison. 

Machine learning models have proven effective in forecasting tasks due to 
their ability to capture complex non-linear relationships in data. They 
are also relatively easy to train and tend to deliver strong performance 
when leveraging cross-learning techniques. In this study, we experimented 
with Linear (Pooled) Regression and four widely used tree-based methods.
While machine learning models are easier to train compared to deep learning 
alternatives, they often rely heavily on high-quality feature engineering 
\citep{fortrees}. For our experiments, we adopted simplified versions of the 
feature engineering pipelines used in the top-performing M4, M5 and VN1 solutions. 
We constructed time series features such as lags, rolling means, expanding 
means, and calendar features (e.g., year, month, week, day of the week). For the
M5 and VN1 datasets, we also included static metadata such as store, product, 
category, and location identifiers. For the M5 dataset, additional 
external features like special events were also incorporated.
Model hyperparameters were selected based on configurations used in 
top-performing competition solutions when available; otherwise, we relied on 
the recommended defaults from the respective automatic libraries. 
We compared five machine learning models for time series forecasting. 
Linear Regression (LR) is a classical statistical model effective with 
appropriate feature engineering, often used as a benchmark in global model 
performance (\citep{princlocglob}, \citep{monashrepo}). 
Random Forest (RF), an ensemble learning method, captures non-linear patterns 
and is robust to overfitting, making it suitable for demand forecasting 
(\citep{randomforest}, \citep{fortrees}). 
Extreme Gradient Boosting (XGBoost) and Light Gradient Boosting Machine (LGBM) 
are gradient-boosted models known for speed and accuracy, with LGBM excelling 
in large, high-dimensional datasets (\citep{xgb}, \citep{lgbm}, \citet{m5acc}). 
Categorical Boosting (CatBoost) specializes in handling categorical data with 
minimal preprocessing \citep{catboost}. 

Deep learning models have gained significant traction in time series 
forecasting due to their ability to model long-range temporal dependencies 
and to learn hierarchical representations directly from raw input data 
\citep{deeplearn}. 
Unlike machine learning models, deep learning architectures typically do 
not require extensive manual feature engineering. They can autonomously 
learn lagged structures, trend, and seasonality directly from the raw 
series. Nevertheless, training deep models is often more complex due to the 
higher number of hyperparameters and their sensitivity to configuration 
choices, which can significantly affect performance \citep{esrnn}. In our 
implementation, we followed competition-proven practices by providing only 
static, calendar, and external covariates, together with historical values, 
as inputs. Top solutions' guidelines were adopted for setting the model 
hyperparameters when available, and otherwise, the recommended
defaults from the respective automatic libraries were used.
We compared five deep learning models for time series forecasting.
Multi-Layer Perceptron (MLP) is a versatile, efficient neural network used 
in many time series forecasting tasks \citep{mlp}, while Recurrent Neural 
Networks (RNNs), particularly LSTMs and GRUs, capture temporal dependencies 
in sequential data (\citep{rnn}, \citep{lstm}). 
Temporal Convolutional Networks (TCN) offer an alternative by using causal 
convolutions to capture long-range dependencies \citep{tcn}. 
Neural Basis Expansion Analysis for Time Series (NBEATS) and Neural 
Hierarchical Interpolation for Time Series (NHITS) are state-of-the-art 
models for time series forecasting, with NBEATS offering interpretable trend 
and seasonality components \citep{nbeats}, and NHITS improving upon NBEATS 
with hierarchical interpolation mechanisms \citep{nhits}.

All models were implemented in Python using Nixtla's framework \citep{nixtla}. 
Specifically, the \textit{statsforecast} library was adopted to train univariate 
statistical models, the \textit{mlforecast} library was used to train the global 
machine learning models, while \textit{neuralforecast} was employed for efficiently 
training the global deep learning models.

For this study, we adopted a cloud computing machine NC6s\_v3 hosted on 
Microsoft Azure, with Linux Ubuntu 24 operating system, 1 Graphical 
Processing Unit, 6 Computing Processing Units, 112GB of memory.


\subsection{Evaluation strategy} \label{s_sec:eval_setup}

Out-of-sample evaluation is a cornerstone of time series forecasting, 
providing a way to test a model's generalization ability beyond the 
training period, an essential step given that future data may diverge 
from historical patterns due to structural breaks, level shifts, or 
unexpected shocks \citep{tseval}. Among the various evaluation strategies, 
the rolling origin evaluation has emerged as the most widely accepted and 
rigorous method \citep{tscrossval}. It respects the chronological nature 
of time series data while enabling repeated assessments across multiple 
forecast origins. In this framework, the time series is split into a 
training set and a test set, where the model is trained on the former and 
evaluated on the latter. Forecasts are generated for a defined horizon \(h\), 
and at each iteration, the forecast origin is shifted forward by a specified 
step size, typically one, to simulate real-time forecasting. The model is then 
retrained on the updated training data, using either a fixed-length or 
expanding window. Evaluation metrics (see Section \ref{s_sec:perf_met}) are 
averaged over all iterations to provide a robust estimate of forecasting 
accuracy.

Compared to fixed origin evaluations, rolling origin offers 
a more nuanced view of a model’s robustness by exposing it to a range of 
temporal conditions (including seasonal patterns, trends, and potential 
anomalies) that may not be captured in a single evaluation window 
\citep{tscrossval}. This iterative approach reduces the risk of overfitting 
to one specific train-test split, and is especially beneficial in operational 
contexts like retail, energy, logistics, or finance, where forecasts are 
regularly updated in response to incoming data. It also aligns well with 
real-world decision-making processes that require forecasts to adapt over time.

In practice, the design of a rolling origin evaluation depends on several 
parameters: the size of the test set, the length of the forecast horizon, the 
step size, and the windowing strategy (fixed vs expanding). The expanding 
window strategy is particularly suited for short time series, as it maximizes 
the amount of historical data available at each iteration. This approach is 
widely used in applied settings \citep{fortheopract}, and it is the one we 
adopted in our study, especially given the relatively short length of the 
weekly VN1 series. 
The Table \ref{tab:retrain_scenario} outlines the key parameters used in our
experiments, including a test set spanning at least one full year, to mitigate
seasonal bias, horizons aligned with typical business use cases, and a step 
size of one to maximize evaluation frequency. A step size of one is also relevant
in the context of stability evaluation, since multiple forecasts for the same 
target are required to compute stability metrics.

\begin{table}[ht]
    \caption{Retraining set, train and test size, frequency, forecast horizon,
    and step size for each dataset. The setup is based on the respective time 
    series frequency.}
    \centering
    \begin{tabular}{lll} 
    \toprule
    & \textbf{M4 / M5} & \textbf{VN1} \\
    \midrule
    Frequency (\(f\)) & daily (7) & weekly (52) \\
    Train size (\(n\)) & $\geq 730$ & $\geq 157$ \\
    Test size (\(T\)) & 364 & 52 \\
    Horizon (\(h\)) & 28  & 13 \\
    Step size (\(p\)) & 1 & 1 \\
    Retrain set (\(r\)) & \{7, 14, 21, 30, 60, 90, 120, 150, 180, 364\} & \{1, 2, 3, 4, 6, 8, 10, 13, 26, 52\} \\
    \bottomrule
    \end{tabular}
    \label{tab:retrain_scenario}
\end{table}

Furthermore, to investigate the impact of retraining on forecasting 
stability, we examined various retraining scenarios, or retraining windows. 
As defined in \citet{zanotti1}, each scenario \(r\) indicates the number of new 
observations after which the model is retrained. These retraining scenarios 
are tailored to the data frequency, daily for M4 and M5, and weekly for VN1, since 
frequency dictates both the forecast horizon and business review cycles. 
For instance, in the daily case, \(r=7\) reflects weekly updates. 
Across all scenarios, we trained global models on each dataset, updating 
the model either completely at each retraining step or not at all. We excluded 
hyperparameter tuning due to its high cost and marginal expected benefit. 
We treated \(r=1\) (or \(r=7\) for daily data) as the accuracy benchmark and 
\(r=T\) as the no-retraining baseline, with intermediate values representing 
periodic retraining strategies.


\subsection{Evaluation metrics} \label{s_sec:perf_met}

\subsubsection{Performance metrics} \label{ss_sec:acc_met}

Evaluating point forecast accuracy in time series is a debated topic, 
with no consensus on the best metric \citep{forevalds}. In our context, 
characterized by intermittent data, metrics based on absolute or percentage 
errors are suboptimal \citep{evalbestacc}, and due to varying scales across 
series, a scaled accuracy metric should be preferred. Hence, we adopted the 
Root Mean Squared Scaled Error (RMSSE) \citep{evalmeasacc},

\begin{equation}
   \text{RMSSE} = \sqrt{\frac{\frac{1}{h} \sum_{t=n+1}^{n+h} 
   (y_t - \hat{y}_t)^2}{\frac{1}{n-s} \sum_{t=s+1}^{n} 
   (y_t - y_{t-s})^2}} \mbox{.}
   \label{eq:rmsse}
\end{equation}

which compares the model’s squared error to that of a seasonal naive 
forecast, and it was the official accuracy metric in the M5 competition
(with \(s=1\)) \citep{m5comp}. Lower values indicate better performance.

To assess probabilistic accuracy, we evaluated the quantiles produced
using a scaled version of the Multi-Quantile Loss. The Scaled Quantile 
Loss, also known as (Scaled) Pinball Loss, and the Scaled Multi-Quantile 
Loss are defined as:

\begin{equation}
    \text{SQL}(\alpha) = \frac{
    \frac{1}{h} \sum_{t = n+1}^{n+h} 
    \Big( 
    \alpha \cdot (y_t - \hat{q}_t^{(\alpha)}) \cdot 
    \mathbb{I}_{\hat{q}_t^{(\alpha)} \leq y_t} 
    + 
    (1 - \alpha) \cdot (\hat{q}_t^{(\alpha)} - y_t) \cdot 
    \mathbb{I}_{\hat{q}_t^{(\alpha)} > y_t} 
    \Big)
    }{\frac{1}{n-s} \sum_{t=s+1}^{n} |y_t - y_{t-s}|} \mbox{,}
    \label{eq:sql}
\end{equation}

\begin{equation}
    \text{SMQL} = 
    \frac{1}{|\mathcal{Q}|} \sum_{\alpha \in \mathcal{Q}} \text{SQL}(\alpha) \mbox{,}
    \label{eq:smql}
\end{equation}

These proper scoring rules assess forecast distribution accuracy 
\citep{evalcountdata}. Moreover, the SMQL was the official metric of 
the M5 Uncertainty competition (weighted, and with \(s=1\)) \citep{m5unc}.

We considered a total of 14 different quantiles of probability levels 
$\alpha \in$ \{0.005, 0.025, 0.050, 0.100, 0.150, 0.200, 0.025, 0.750, 
0.800, 0.850, 0.900, 0.950, 0.975, 0.995\}. Therefore, we analyzed the 50\% 
60\%, 70\%, 80\%, 90\%, 95\%, and 99\% central prediction intervals.
The 50\% and 60\% central prediction intervals provide a good description of 
the center of the forecast distribution, while the 90\%, 95\%, and 99\% give more 
information about its tails. 
These quantiles provide sufficient information about the uncertainty
of the forecasts and allow for the effective description of the whole distribution.
To ensure reliable quantile estimates, conformal prediction intervals were 
computed on a validation set at least twice the forecast horizon, which 
constrained the number of series used 
\footnote{
    We used a validation set four times larger than the forecast horizon for
    the M4 and M5 datasets, and twice the forecast horizon for the VN1 dataset. 
    This limited the number of time series from 4,227 to 2,557 for the M4 
    dataset, and from 30,490 to 28,298 for the M5 dataset.
}.

We also normalized each evaluation metric relative to the baseline retraining 
scenario, defined by the dataset frequency, to enable consistent comparison 
across models and retraining periods. To statistically validate our 
findings, we applied the Friedman-Nemenyi test \citep{testfriednem}.

\subsubsection{Stability metrics} \label{ss_sec:stab_met}

The stability evaluation of point predictive models is a crucial 
topic in time series forecasting: very few metrics are available 
to capture models' vertical stability, and there is no consensus 
in the literature on what metric to use \citep{stab1}. \citet{stab2} 
proposed the symmetric Mean Absolute Percentage Change (sMAPC), 
which measures the change of one to \(h\)-step ahead forecasts 
obtained by two consecutive forecast origins, providing a 
measurement of up to which extent the forecasts generated at the 
first origin are unstable compared to those generated at the 
second origin.

\begin{equation}
   \text{sMAPC} = \frac{200}{h - 1} \sum_{t=n+1}^{n+h-1} 
   \frac{\lvert \hat{y}_{t,n} - \hat{y}_{t,n-1} \rvert}
   {\lvert \hat{y}_{t,n} \rvert - \lvert \hat{y}_{t,n-1} \rvert} 
   \mbox{.}
   \label{eq:sMAPC}
\end{equation}
Here, \(\hat{y}_{t,n}\) and \(\hat{y}_{t,n-1}\) are the forecasts 
generated for period \(t\) with origins \(n\) and \(n-1\) respectively.
The instability across different pairs of consecutive forecasting 
origins can be obtained by a simple average of the sMAPC values.
Lower values imply more stable predictions.

Beyond point forecasts, our study also emphasizes the probabilistic 
stability of the models under different retraining strategies. 
Since most machine learning and deep learning models do not natively 
produce full predictive distributions, we employed Conformal Inference 
to construct prediction intervals around point forecasts. Its advantages 
(validity guarantees, minimal assumptions, computational simplicity, and 
effectiveness with limited data) make it especially suitable for our global
forecasting setup, where different models are compared across heterogeneous 
datasets and varying retraining scenarios
\footnote{
    One may prefer, instead, to train models using quantile losses to directly 
    yield quantile forecasts, or to use inherently probabilistic models.
    These are also valid ways of modeling uncertainty, as demonstrated by the 
    M5 Uncertainty competition \citep{m5unc, m5uncsol1}. However, comparing 
    different probabilistic modeling frameworks is out of the scope of our
    analysis, and it may be an area of future research.
}.
Because no metric exists to evaluate the vertical stability (or instability) 
of the resulting probabilistic quantile forecasts, we proposed a new measure, 
the Multi-Quantile Change (MQC), to comprehensively assess the stability of
the probabilistic predictions. We defined the Quantile Change and the 
Multi-Quantile Change as:

\begin{equation}
    \text{QC}(\alpha) = \frac{1}{h-1} \sum_{t=n+1}^{n+h-1} 
    \Big( 
    \alpha \cdot (\hat{q}_{t,n-1}^{(\alpha)} - \hat{q}_{t,n}^{(\alpha)}) \cdot 
    \mathbb{I}_{\hat{q}_{t,n}^{(\alpha)} \leq \hat{q}_{t,n-1}^{(\alpha)}} 
    + (1 - \alpha) \cdot (\hat{q}_{t,n}^{(\alpha)} - \hat{q}_{t,n-1}^{(\alpha)}) \cdot 
    \mathbb{I}_{\hat{q}_{t,n}^{(\alpha)} > \hat{q}_{t,n-1}^{(\alpha)}} 
    \Big) 
    \mbox{,}
    \label{eq:qc}
\end{equation}

\begin{equation}
    \text{MQC} = 
    \frac{1}{|\mathcal{Q}|} \sum_{\alpha \in \mathcal{Q}} \text{QC}(\alpha) \mbox{,}
    \label{eq:mqc}
\end{equation}
where \(\mathcal{Q}\) is the set of possible quantile levels \(\alpha\).
The Quantile Change (QC) metric is a measure of the change in forecasted 
quantiles, $\hat{q}_t^{(\alpha)}$, between two consecutive forecast origins. 
Given two forecast origins, \(n\) and \(n-1\), QC quantifies the average 
adjustment in the predicted quantiles across the forecast horizon \(h\). 
The formula mirrors the Quantile Loss but replaces the actual 
observations \(y_{t}\), with previous quantile forecasts \(\hat{q}_{t,n-1}^{(\alpha)}\), 
thereby shifting the focus from forecast accuracy to forecast stability.
The intuition is straightforward: if the forecasts are stable across time 
(i.e., \(\hat{q}_{t,n}^{(\alpha)} \approx \hat{q}_{t,n-1}^{(\alpha)}\)), then QC 
will be small. Large QC values indicate that the model updates its distribution 
significantly with each new observation, suggesting a lack of stability in 
its probabilistic output.
By aggregating this across a set of quantile levels \(\mathcal{Q}\), the 
Multi-Quantile Change (MQC) captures overall instability, providing a 
comprehensive metric for probabilistic stability.

One drawback of the MQC is that it is a scale-dependent measure 
of instability. To solve this issue, we also propose a scaled version of 
the MQC, the Scaled Multi-Quantile Change (SMQC), defined as

\begin{equation}
    \text{SQC}(\alpha) =
    \frac{
    \frac{1}{h-1}
    \sum_{t=n+1}^{n+h-1} 
    \Big( 
    \alpha \cdot (\hat{q}_{t,n-1}^{(\alpha)} - \hat{q}_{t,n}^{(\alpha)}) \cdot 
    \mathbb{I}_{\hat{q}_{t,n}^{(\alpha)} \leq \hat{q}_{t,n-1}^{(\alpha)}} 
    + (1 - \alpha) \cdot (\hat{q}_{t,n}^{(\alpha)} - \hat{q}_{t,n-1}^{(\alpha)}) \cdot 
    \mathbb{I}_{\hat{q}_{t,n}^{(\alpha)} > \hat{q}_{t,n-1}^{(\alpha)}} 
    \Big) 
        }{
            \frac{1}{n-s} \sum_{t=s+1}^{n} \left| y_t - y_{t-s} \right|
        }
     \mbox{,}
    \label{eq:sqc}
\end{equation}

\begin{equation}
    \text{SMQC} = 
    \frac{1}{|\mathcal{Q}|} \sum_{\alpha \in \mathcal{Q}} \text{SQC}(\alpha) \mbox{,}
    \label{eq:smqc}
\end{equation}
where we adopted the in-sample one-step-ahead seasonal naive forecast as a scaling 
factor, to enable the effective application of this metric to forecasting models 
by ensuring that the results are not biased by the scale of individual time series.

To further contextualize the proposed metric, it is important to compare it 
with existing distributional distance metrics such as the Kullback–Leibler ($KL$) 
divergence and the Wasserstein distance ($W_2$). KL divergence measures the 
relative entropy between two probability distributions and is particularly 
sensitive to differences in the tails \citep{kl}. However, it requires access 
to full predictive densities, which can limit its use in forecasting applications 
where outputs are often expressed only as quantiles. Wasserstein distance ($W_2$), 
by contrast, quantifies the minimal cost of transforming one distribution into 
another and effectively captures both location and shape differences, but it 
usually relies on full distributions (although it can be approximated through 
discretization) \citep{wass}. 
The MQC metric addresses a complementary objective: it evaluates the stability 
of probabilistic forecasts directly at the quantile level, without the need for 
full predictive densities. 
Since it is based solely on a finite set of quantiles rather than the full 
predictive distribution, MQC does not capture all aspects of the distributional 
change, potentially missing more subtle shifts in shape, skewness, or 
variance that occur between those quantiles. However, in practical 
forecasting applications, especially in retail and supply chain domains 
where decisions are often made based on specific quantile levels (e.g., 
high quantiles for safety stock), this limitation is mitigated by the 
fact that the selected quantiles are usually those of greatest operational 
interest.
Importantly, MQC is both model-agnostic and distribution-free, meaning that it 
can be applied consistently across different forecasting approaches without 
requiring strong assumptions about the underlying predictive distributions.
Moreover, MQC is computationally lightweight compared to the Wasserstein distance 
or the KL divergence, which makes it suitable for large-scale applications 
involving thousands of time series.


\section{Results and discussion} \label{sec:results}

This section discusses the empirical findings of our study, highlighting 
the interplay between accuracy and stability across different retraining 
strategies. 

\begin{figure}
    \centering
    \includegraphics[scale=0.65]{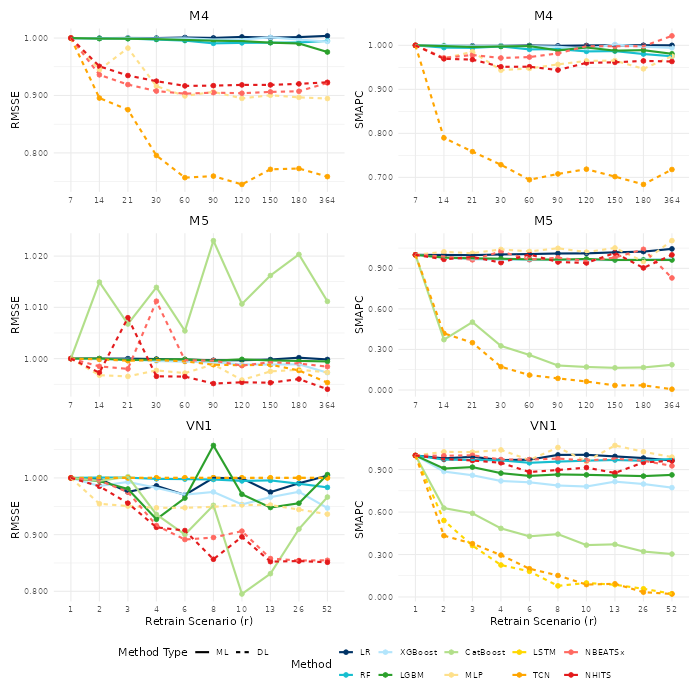}
    \caption{RMSSE and sMAPC results for each global method and 
    retrain scenario combination in relative terms 
    with respect to the baseline scenario, \(r = 7\) 
    for the M4 and M5 datasets, and \(r = 1\) for the VN1 dataset.}
    \label{fig:global_rmsse_smapc}
\end{figure} 

Figure \ref{fig:global_rmsse_smapc} shows the point forecast accuracy (on
the left) and the point forecast stability (on the right) of each model 
\footnote{
    For the long computing time required, the Random Forest model has not 
    been trained on the M5 dataset, while the LSTM model has not been adopted
    on both the M4 and M5 datasets. Moreover, the CatBoost method has not been 
    used on the M4 dataset since too few categorical variables are available 
    for this dataset.
}
along the different retraining scenarios for the M4, M5, and VN1 datasets.
To allow comparisons, we display the results in relative terms with respect 
to the baseline scenarios, that is \(r = 7\) for M4 and M5, and \(r = 1\) for 
VN1.
In terms of forecast stability, the models' behavior is similar across
the different datasets. As expected, the sMAPC profiles are non-increasing 
functions of the retraining scenario. The stability remains practically the 
same, or even improves, as the retraining period increases. 
Indeed, moving from a high to a low retraining scenario, the stability of 
most global models improves compared to the baseline.
In particular, for some models (i.e. CatBoost, LSTM, and TCN), the stability
improvement is strong and consistent, reaching very low levels. 
These results imply that less frequent retraining may have a huge impact on 
the point forecast stability of global models. This can be explained by the 
fact that, when a forecasting origin is updated, and the model is retrained on 
the new data, the model's predictions may differ consistently (even using the 
cross-learning approach) from those produced before the update. Therefore, 
less retraining is often synonymous with more stability. 
The statistical tests (see the Supplementary material) confirm the above 
results, showing that most periodic retraining scenarios are statistically 
different to the continuous retraining strategy in terms of point forecast 
stability at the 5\% level.
Similar considerations arise for the point forecast accuracy: the RMSSE 
profiles of the global models remain remarkably stable across retraining 
scenarios, sometimes displaying even modest improvements in accuracy when 
retrained less frequently (as is the case of deep learning models for the
M4 dataset). In particular, under low periodic retraining schemes, the 
performance of most global models is statistically indistinguishable from 
the baseline configuration, and even under extreme settings (such as the 
no-retraining scenario) any degradation in accuracy remains below 5\%.
Overall, these findings show that global models can preserve, and in 
some cases even enhance, their forecasting accuracy over extended 
retraining intervals. Hence, frequent retraining may introduce unnecessary 
variability without delivering tangible gains in both predictive 
performance and forecast stability. 

\begin{figure}
    \centering
    \includegraphics[scale=0.65]{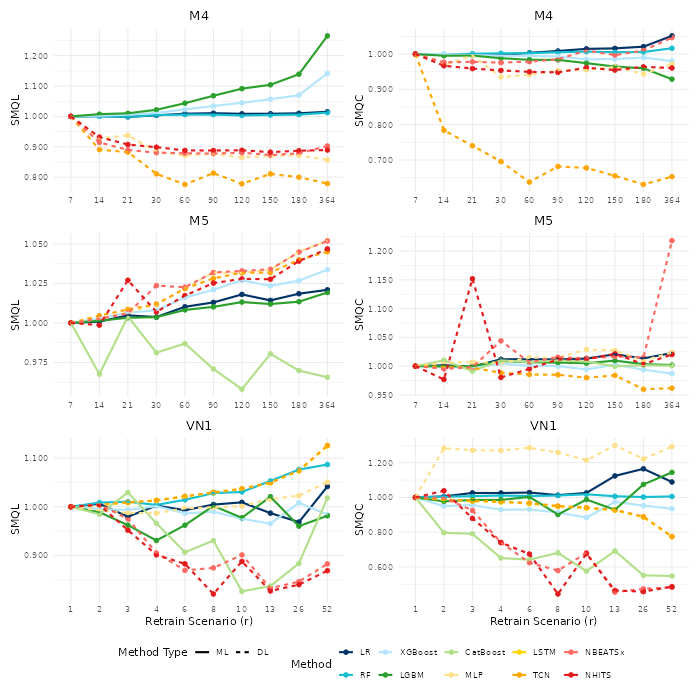}
    \caption{SMQL and SMQC results for each method and 
    retrain scenario combination in relative terms 
    with respect to the baseline scenario, \(r = 7\) 
    for the M4 and M5 datasets, and \(r = 1\) for the VN1 dataset.}
    \label{fig:global_smql_smqc}
\end{figure} 

In a similar fashion, Figure \ref{fig:global_smql_smqc} summarizes the 
relative performance (on the left) and stability (on the right) from a 
probabilistic forecasting perspective. 
In this context, we observe that, for daily data, the stability 
paths are constant across almost every retraining scenario, meaning 
that less frequent updates neither harm nor improve the probabilistic 
forecasting stability of most global models
\footnote{
    There are a few spikes in the SMQC profiles of some models. 
    These can be due to different factors (e.g., inappropriate 
    hyperparameter values). Nevertheless, the profiles are stable 
    overall.
}. 
For weekly data, instead, we observe some differences among the models.
Some of them show constant improvements in stability as the retraining 
scenario increases, while others do not exhibit particular enhancements 
or deteriorations in most periodic retraining intervals. 
When shifting the focus from point to probabilistic forecasting, the 
impact of retraining frequency on accuracy becomes more pronounced. Unlike 
point forecasts, probabilistic accuracy is more sensitive to longer 
retraining intervals. In several cases, reducing the retraining frequency 
leads to a gradual deterioration in probabilistic performance, indicating 
that the estimation of predictive uncertainty is more affected by 
outdated model parameters than point predictions. Nevertheless, these 
effects remain limited for low periodic retraining scenarios, where 
differences in probabilistic accuracy are typically negligible (below 2\%). 
Importantly, this sensitivity is not universal: some models exhibit clear 
accuracy losses, while others maintain stable probabilistic performance or 
even improve. Overall, these findings suggest that accuracy uncertainty 
estimates benefit more from regular updates, yet do not strictly require 
very frequent retraining to remain reliable.

\begin{figure}
    \centering
    \includegraphics[scale=0.65]{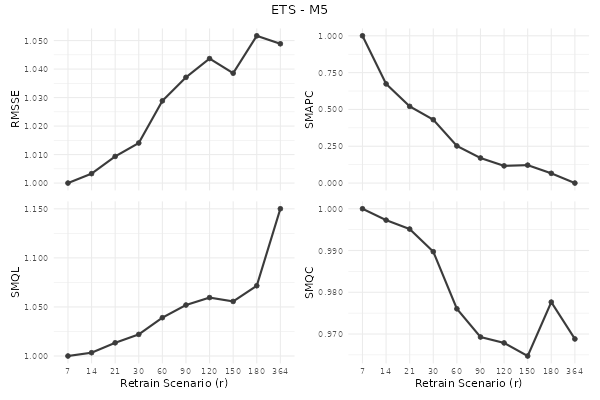}
    \caption{RMSSE, sMAPC, SMQL, and SMQC results for ETS method 
    on the M5 dataset for each retrain scenario combination in 
    relative terms with respect to the baseline scenario, \(r = 7\).}
    \label{fig:ets_m5}
\end{figure} 

To further contextualize the stability findings for global models, we also 
evaluated two classical local forecasting approaches: automatic ARIMA and 
automatic ETS
\footnote{
    For the long computing time required, the ARIMA model has not been 
    trained on the M5 dataset.
}. 
Both local models exhibit highly consistent behaviors across 
datasets and retraining scenarios (see Supplementary material). As an 
illustrative example, Figure \ref{fig:ets_m5} reports the results for ETS on 
the M5 dataset. The accuracy profiles for both point and probabilistic 
forecasting, measured by RMSSE and SMQL respectively, are strictly increasing 
functions of the retraining period, indicating a systematic deterioration 
in performance as retraining becomes less frequent. Conversely, as expected, 
the stability profiles, captured by sMAPC and SMQC, are strictly decreasing, 
showing that less frequent retraining leads to progressively more stable 
forecasts. 
This monotonic trade-off highlights an intrinsic property of local models: 
stability improvements are usually obtained at the cost of accuracy. 
In contrast, global models display markedly different dynamics, often 
maintaining both accuracy and stability over a wide range of retraining 
periods, and, in some cases, even improving along one dimension without 
compromising the other. This comparison suggests that global forecasting 
models are inherently more robust to reduced retraining, benefiting from 
cross-series learning that mitigates the sensitivity to individual time 
series updates, a property that local models, by construction, cannot exploit.

Overall, the results from Figures \ref{fig:global_rmsse_smapc} and 
\ref{fig:global_smql_smqc}, supported by the statistical tests, 
indicate that reducing the retraining frequency of global models does not 
negatively impact (and may even enhance) forecast stability, for both point 
and probabilistic forecasts. These results indicate that the accuracy–stability 
trade-off in global forecasting models is substantially milder than in local 
approaches. 
When considered alongside the findings of \citet{zanotti1} on energy 
consumption, this provides strong evidence against the practice of frequent 
retraining in global forecasting models. Lower retraining scenarios 
preserve both accuracy and stability, while offering substantial savings 
in computational resources and energy consumption of forecasting systems.

\begin{figure}
    \centering
    \includegraphics[scale=0.65]{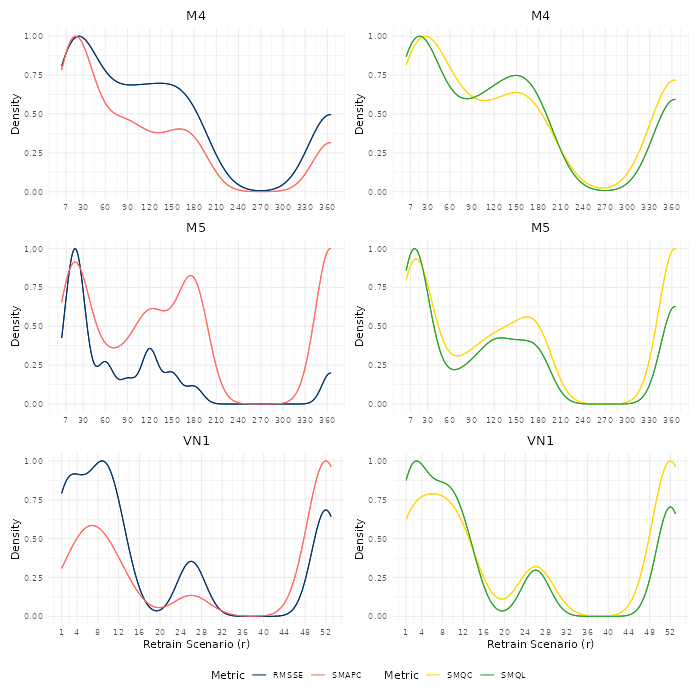}
    \caption{The optimal retraining frequency identified for each 
    forecasting objective, and for each series in the datasets 
    considering all the models.}
    \label{fig:optimal_freq}
\end{figure} 

Finally, to jointly assess how accuracy- and stability-driven objectives 
translate into different retraining policies, Figure \ref{fig:optimal_freq} 
provides a distributional view of the optimal retraining frequency across 
time series, in both point and probabilistic forecasting settings.
Overall, the results indicate that continuous retraining is not required to 
achieve either accurate or stable forecasts when using global models. 
Across both point and probabilistic objectives, the retraining frequency can 
be safely reduced in favor of a periodic strategy, with stability-based 
criteria generally favoring slightly less frequent updates, in some cases 
even no retraining. However, compared to point forecasting, probabilistic 
objectives tend to favor slightly more frequent retraining, suggesting that 
accurate and stable uncertainty estimation requires a more regular model 
re-estimation. In practical terms, for daily data, a retraining 
interval of approximately 21–30 days emerges as a reasonable compromise 
across objectives, while for weekly data, retraining can be postponed 
usually up to 8 weeks without compromising performance. 
These retraining schedules strike a balance between the need to update models 
in response to potential distributional changes and the desire to limit 
unnecessary updates. Importantly, adopting such periodic retraining strategies 
can reduce the computational cost of forecasting by up to 75\% \citet{zanotti1}, 
leading to more efficient and sustainable forecasting systems without 
sacrificing accuracy and also improving forecast stability.


\section{Conclusions} \label{sec:conc}

This study investigates the role of retraining frequency in global 
forecasting models, with a particular focus on its implications for 
both forecast accuracy and forecast stability in point and probabilistic 
settings. Using three large-scale benchmark datasets (M4, M5, and VN1), 
we provided an extensive empirical analysis across multiple machine 
learning and deep learning architectures, complemented by a comparison 
with classical local forecasting methods. These results contribute to 
a more nuanced understanding of retraining as a design choice in 
large-scale forecasting systems, rather than as a default operational 
requirement.

Across all datasets and modeling families, we find that reducing the 
retraining frequency does not harm and often improves forecast stability. 
This result holds for both point and probabilistic forecasts, as 
measured by the proposed Scaled Multi-Quantile Change (SMQC) metric. 
In particular, stability profiles are typically flat or monotonically 
improving as retraining becomes less frequent, indicating that frequent 
retraining often introduces unnecessary variability in forecasts 
without delivering tangible benefits. These results are statistically 
validated via statistical tests and indicate that continuous retraining 
is often not the optimal strategy.

While point forecast accuracy is largely unaffected by reduced retraining,
remaining stable or even improving in many cases, probabilistic accuracy 
shows a slightly higher sensitivity to retraining frequency. Less frequent 
updates may lead to modest degradations in distributional accuracy, 
although these effects are generally negligible for low periodic retraining 
scenarios and remain limited even under no-retraining regimes. 
Contrary to the common view of a strict trade-off between forecasting 
accuracy and stability, our results indicate that, in the context of 
global forecasting models, this tension is largely attenuated. Moreover, 
in many cases, reducing the retraining frequency does not require sacrificing 
predictive performance and may even contribute to both more stable and more 
accurate forecasts. This behavior contrasts with what is often observed in 
local modeling settings, where model re-estimation can lead to pronounced 
fluctuations in both accuracy and forecast revisions. For global models, 
the shared learning across series appears to act as a regularizing 
mechanism, making them less sensitive to frequent updates and allowing 
lower retraining frequencies to jointly improve stability while preserving, 
or even enhancing, also accuracy.

From a practical standpoint, these findings have several implications. 
First, businesses that rely on large-scale forecasting systems, such as 
those in retail, energy, and supply chain planning can adopt lower 
retraining strategies not only to reduce computational resources and 
the associated costs of forecasting systems, but also to increase 
operational stability, without compromising on forecast accuracy. 
Indeed, our results strongly support low, periodic retraining strategies
(i.e. every 30 days or every 8 weeks) in global forecasting systems. 
Finally, these results promote a shift in emphasis from pure accuracy 
to stability-aware model selection and evaluation, especially in those 
contexts that are more sensitive to forecast reviews.

Nevertheless, some limitations of the present study should be acknowledged. 
While the analysis covers a broad range of global models and three 
state-of-the-art forecasting datasets, exploring the effects of 
retraining newer methods and different datasets may be a concrete area
for future research. 
Moreover, this work focuses on the stability of global models. A deeper 
analysis of local approaches could therefore represent a valuable direction 
for future research.
Additionally, the study focuses solely on full retraining schemes and does 
not explore incremental or online learning strategies, which may offer a 
different trade-off between accuracy and stability. Future research could 
extend this work by incorporating such adaptive methods and introducing 
business-oriented stability metrics that link forecast stability directly 
to operational cost or risk. 
Future research should also investigate the interplay between retraining 
frequency and hyperparameter optimization. Different models may exhibit 
heterogeneous responses to retraining strategies, depending on their 
hyperparameter configurations, and a deeper understanding of these 
interactions could enable the design of even more efficient and robust 
forecasting strategies.
Furthermore, this study employed a single framework for generating 
probabilistic forecasts, namely Conformal Inference. Although conformalized 
residuals provide a valid means for quantifying uncertainty, future 
research could address this limitation by studying probabilistic stability 
in native probabilistic models, thereby enhancing the robustness and 
generalizability of the findings.
Finally, further theoretical development of stability measures, particularly 
for probabilistic forecasts, may also contribute to a deeper understanding 
of how models behave across time and under varying data regimes.

In summary, this work provides strong empirical evidence that frequent 
retraining is neither necessary nor desirable in many forecasting 
applications. By explicitly accounting for stability alongside accuracy, 
forecasting systems can be designed to be more robust and computationally 
sustainable.


\bibliography{main}


\newpage
\section*{Supplementary material} \label{sec:supmat}

\begin{table}[ht]
    \centering
    \begin{tabular}{lcccc}
    \toprule
    \textbf{Model} & \textbf{Learning rate} & \textbf{Iterations} & \textbf{Feature fraction} \\
    \midrule
    Random Forest & 0.03 & 1200 & 0.85 \\
    XGBoost       & 0.03 & 1200 & 0.85 \\
    LightGBM      & 0.03 & 1200 & 0.85 \\
    CatBoost      & 0.03 & 1200 & --- \\
    \bottomrule
    \end{tabular}
    \caption{Main hyperparameter settings for machine learning forecasting models.}
    \label{tab:hyper_ml}
\end{table}

\begin{table}
    \centering
    \begin{tabular}{lcccc}
    \toprule
    \textbf{Model} & \textbf{Learning rate} & \textbf{Epochs} & \textbf{Scaler} \\
    \midrule
    MLP     & 0.001 & 1000 & Robust \\
    LSTM    & 0.001 & 1000 & Robust \\
    TCN     & 0.001 & 1000 & Robust \\
    NBEATS  & 0.001 & 1000 & Robust \\
    NHITS   & 0.001 & 1000 & Robust \\
    \bottomrule
    \end{tabular}
    \caption{Main hyperparameter settings for deep learning forecasting models.}
    \label{tab:hyper_dl}
\end{table}

\begin{table}
    \centering
    \begin{tabular}{lrrrrrrrrrr}
    \toprule
    Method & 7 & 14 & 21 & 30 & 60 & 90 & 120 & 150 & 180 & 364 \\ 
    \midrule
      ETS & 1.606 & 1.760 & 1.925 & 2.035 & 2.510 & 3.041 & 3.341 & 3.704 & 4.113 & 5.449 \\ 
      ARIMA & 1.621 & 1.776 & 1.944 & 2.056 & 2.545 & 3.094 & 3.356 & 3.804 & 4.276 & 5.674 \\ 
      LR & 1.578 & 1.578 & 1.578 & 1.578 & 1.580 & 1.578 & 1.581 & 1.579 & 1.581 & 1.584 \\ 
      RF & 1.715 & 1.713 & 1.713 & 1.710 & 1.707 & 1.699 & 1.700 & 1.701 & 1.701 & 1.706 \\ 
      XGBoost & 18.125 & 18.115 & 18.140 & 18.142 & 18.144 & 18.116 & 18.130 & 18.026 & 18.088 & 18.112 \\ 
      LGBM & 15.467 & 15.452 & 15.450 & 15.435 & 15.413 & 15.393 & 15.386 & 15.282 & 15.261 & 15.034 \\ 
      MLP & 1.705 & 1.611 & 1.676 & 1.562 & 1.533 & 1.546 & 1.526 & 1.536 & 1.529 & 1.526 \\ 
      TCN & 2.280 & 2.043 & 1.997 & 1.815 & 1.727 & 1.733 & 1.699 & 1.758 & 1.761 & 1.730 \\ 
      NBEATSx & 1.625 & 1.521 & 1.494 & 1.475 & 1.468 & 1.471 & 1.469 & 1.473 & 1.475 & 1.498 \\ 
      NHITS & 1.629 & 1.548 & 1.522 & 1.506 & 1.493 & 1.494 & 1.496 & 1.496 & 1.499 & 1.503 \\ 
    \bottomrule
    \end{tabular}
    \caption{M4 RMSSE values for each method and retrain scenario combination.}
    \label{tab:m4_rmsse}
\end{table}

\begin{table}
    \centering
    \begin{tabular}{lrrrrrrrrrr}
    \toprule
    Method & 7 & 14 & 21 & 30 & 60 & 90 & 120 & 150 & 180 & 364 \\ 
    \midrule
      ETS & 0.014 & 0.010 & 0.008 & 0.007 & 0.005 & 0.004 & 0.003 & 0.003 & 0.002 & 0.000 \\ 
      ARIMA & 0.014 & 0.010 & 0.008 & 0.007 & 0.005 & 0.004 & 0.003 & 0.003 & 0.002 & 0.000 \\ 
      LR & 0.013 & 0.013 & 0.013 & 0.013 & 0.013 & 0.013 & 0.013 & 0.013 & 0.013 & 0.013 \\ 
      RF & 0.013 & 0.013 & 0.013 & 0.013 & 0.013 & 0.013 & 0.013 & 0.013 & 0.013 & 0.013 \\ 
      XGBoost & 0.052 & 0.052 & 0.052 & 0.052 & 0.052 & 0.052 & 0.052 & 0.052 & 0.052 & 0.052 \\ 
      LGBM & 0.123 & 0.123 & 0.123 & 0.123 & 0.123 & 0.122 & 0.122 & 0.122 & 0.122 & 0.121 \\ 
      MLP & 0.017 & 0.016 & 0.017 & 0.016 & 0.016 & 0.016 & 0.016 & 0.016 & 0.016 & 0.016 \\ 
      TCN & 0.021 & 0.017 & 0.016 & 0.015 & 0.015 & 0.015 & 0.015 & 0.015 & 0.014 & 0.015 \\ 
      NBEATSx & 0.015 & 0.015 & 0.015 & 0.015 & 0.015 & 0.015 & 0.015 & 0.015 & 0.015 & 0.016 \\ 
      NHITS & 0.016 & 0.015 & 0.015 & 0.015 & 0.015 & 0.015 & 0.015 & 0.015 & 0.015 & 0.015 \\ 
    \bottomrule
    \end{tabular}
    \caption{M4 sMAPC values for each method and retrain scenario combination.}
    \label{tab:m4_smapc}
\end{table}

\begin{table}
    \centering
    \begin{tabular}{lrrrrrrrrrr}
    \toprule
    Method & 7 & 14 & 21 & 30 & 60 & 90 & 120 & 150 & 180 & 364 \\ 
    \midrule
      ETS & 0.419 & 0.467 & 0.513 & 0.549 & 0.684 & 0.864 & 0.948 & 1.078 & 1.230 & 1.563 \\ 
      ARIMA & 0.428 & 0.475 & 0.524 & 0.560 & 0.701 & 0.881 & 0.955 & 1.111 & 1.276 & 1.655 \\ 
      LR & 0.394 & 0.394 & 0.393 & 0.395 & 0.397 & 0.398 & 0.397 & 0.397 & 0.398 & 0.400 \\ 
      RF & 0.437 & 0.438 & 0.437 & 0.439 & 0.440 & 0.440 & 0.439 & 0.439 & 0.440 & 0.443 \\ 
      XGBoost & 2.279 & 2.288 & 2.293 & 2.309 & 2.337 & 2.365 & 2.394 & 2.405 & 2.447 & 2.636 \\ 
      LGBM & 2.455 & 2.473 & 2.480 & 2.511 & 2.564 & 2.624 & 2.680 & 2.707 & 2.793 & 3.096 \\ 
      MLP & 0.501 & 0.465 & 0.470 & 0.443 & 0.437 & 0.439 & 0.433 & 0.436 & 0.437 & 0.429 \\ 
      TCN & 0.652 & 0.583 & 0.577 & 0.531 & 0.508 & 0.532 & 0.506 & 0.528 & 0.521 & 0.507 \\ 
      NBEATSx & 0.466 & 0.426 & 0.415 & 0.410 & 0.409 & 0.409 & 0.410 & 0.407 & 0.409 & 0.421 \\ 
      NHITS & 0.466 & 0.435 & 0.423 & 0.419 & 0.414 & 0.414 & 0.414 & 0.412 & 0.414 & 0.415 \\  
    \bottomrule
    \end{tabular}
    \caption{M4 SMQL values for each method and retrain scenario combination.}
    \label{tab:m4_smql}
\end{table}

\begin{table}
    \centering
    \begin{tabular}{lrrrrrrrrrr}
    \toprule
    Method & 7 & 14 & 21 & 30 & 60 & 90 & 120 & 150 & 180 & 364 \\ 
    \midrule
      ETS & 0.278 & 0.300 & 0.326 & 0.351 & 0.396 & 0.435 & 0.449 & 0.469 & 0.493 & 0.603 \\ 
      ARIMA & 0.279 & 0.301 & 0.328 & 0.347 & 0.397 & 0.439 & 0.451 & 0.482 & 0.496 & 0.593 \\ 
      LR & 0.262 & 0.262 & 0.262 & 0.261 & 0.263 & 0.264 & 0.266 & 0.266 & 0.268 & 0.276 \\ 
      RF & 0.351 & 0.351 & 0.351 & 0.352 & 0.352 & 0.353 & 0.354 & 0.353 & 0.353 & 0.357 \\ 
      XGBoost & 2.156 & 2.150 & 2.154 & 2.153 & 2.149 & 2.142 & 2.134 & 2.126 & 2.130 & 2.119 \\ 
      LGBM & 1.935 & 1.927 & 1.928 & 1.915 & 1.906 & 1.904 & 1.890 & 1.864 & 1.861 & 1.795 \\ 
      MLP & 0.371 & 0.359 & 0.369 & 0.346 & 0.349 & 0.353 & 0.354 & 0.357 & 0.350 & 0.360 \\ 
      TCN & 0.506 & 0.397 & 0.375 & 0.352 & 0.323 & 0.345 & 0.343 & 0.332 & 0.320 & 0.331 \\ 
      NBEATSx & 0.338 & 0.330 & 0.330 & 0.330 & 0.331 & 0.333 & 0.341 & 0.337 & 0.341 & 0.354 \\ 
      NHITS & 0.344 & 0.332 & 0.329 & 0.328 & 0.326 & 0.326 & 0.330 & 0.328 & 0.331 & 0.330 \\  
    \bottomrule
    \end{tabular}
    \caption{M4 SMQC values for each method and retrain scenario combination.}
    \label{tab:m4_smqc}
\end{table}


\begin{table}
    \centering
    \begin{tabular}{lrrrrrrrrrr}
    \toprule
    Method & 7 & 14 & 21 & 30 & 60 & 90 & 120 & 150 & 180 & 364 \\ 
    \midrule
      ETS & 0.747 & 0.751 & 0.756 & 0.759 & 0.769 & 0.776 & 0.778 & 0.774 & 0.784 & 0.782 \\ 
      LR & 0.759 & 0.759 & 0.759 & 0.759 & 0.758 & 0.758 & 0.758 & 0.758 & 0.759 & 0.758 \\ 
      XGBoost & 0.737 & 0.737 & 0.737 & 0.737 & 0.737 & 0.737 & 0.737 & 0.737 & 0.736 & 0.735 \\ 
      LGBM & 0.753 & 0.753 & 0.753 & 0.753 & 0.753 & 0.753 & 0.753 & 0.753 & 0.753 & 0.753 \\ 
      CatBoost & 0.917 & 0.929 & 0.924 & 0.928 & 0.921 & 0.936 & 0.925 & 0.930 & 0.934 & 0.926 \\ 
      MLP & 0.803 & 0.800 & 0.800 & 0.801 & 0.801 & 0.802 & 0.799 & 0.801 & 0.801 & 0.800 \\ 
      TCN & 0.844 & 0.844 & 0.844 & 0.844 & 0.844 & 0.844 & 0.844 & 0.843 & 0.843 & 0.841 \\ 
      NBEATSx & 0.797 & 0.796 & 0.796 & 0.807 & 0.797 & 0.797 & 0.797 & 0.797 & 0.797 & 0.796 \\ 
      NHITS & 0.811 & 0.808 & 0.817 & 0.808 & 0.808 & 0.806 & 0.807 & 0.807 & 0.807 & 0.805 \\ 
    \bottomrule
    \end{tabular}
    \caption{M5 RMSSE values for each method and retrain scenario combination.}
    \label{tab:m5_rmsse}
\end{table}

\begin{table}
    \centering
    \begin{tabular}{lrrrrrrrrrr}
    \toprule
    Method & 7 & 14 & 21 & 30 & 60 & 90 & 120 & 150 & 180 & 364 \\ 
    \midrule
      ETS & 0.104 & 0.070 & 0.054 & 0.045 & 0.026 & 0.018 & 0.012 & 0.013 & 0.007 & 0.000 \\ 
      LR & 0.188 & 0.188 & 0.188 & 0.188 & 0.189 & 0.190 & 0.190 & 0.191 & 0.192 & 0.196 \\ 
      XGBoost & 0.075 & 0.074 & 0.073 & 0.073 & 0.072 & 0.073 & 0.072 & 0.072 & 0.073 & 0.073 \\ 
      LGBM & 0.054 & 0.053 & 0.053 & 0.053 & 0.052 & 0.052 & 0.052 & 0.052 & 0.052 & 0.052 \\ 
      CatBoost & 0.120 & 0.045 & 0.060 & 0.039 & 0.031 & 0.022 & 0.021 & 0.020 & 0.020 & 0.023 \\ 
      MLP & 0.499 & 0.511 & 0.505 & 0.519 & 0.512 & 0.524 & 0.510 & 0.525 & 0.475 & 0.552 \\ 
      TCN & 0.497 & 0.208 & 0.174 & 0.086 & 0.055 & 0.043 & 0.031 & 0.017 & 0.017 & 0.003 \\ 
      NBEATSx & 0.547 & 0.540 & 0.527 & 0.560 & 0.528 & 0.535 & 0.522 & 0.536 & 0.546 & 0.424 \\ 
      NHITS & 0.566 & 0.547 & 0.557 & 0.533 & 0.566 & 0.536 & 0.532 & 0.548 & 0.511 & 0.565 \\  
    \bottomrule
    \end{tabular}
    \caption{M5 sMAPC values for each method and retrain scenario combination.}
    \label{tab:m5_smapc}
\end{table}

\begin{table}
    \centering
    \begin{tabular}{lrrrrrrrrrr}
    \toprule
    Method & 7 & 14 & 21 & 30 & 60 & 90 & 120 & 150 & 180 & 364 \\ 
    \midrule
      ETS & 0.249 & 0.251 & 0.254 & 0.256 & 0.260 & 0.263 & 0.264 & 0.264 & 0.267 & 0.288 \\ 
      LR & 0.267 & 0.268 & 0.269 & 0.268 & 0.270 & 0.271 & 0.272 & 0.271 & 0.272 & 0.273 \\ 
      XGBoost & 0.245 & 0.246 & 0.247 & 0.247 & 0.249 & 0.250 & 0.252 & 0.251 & 0.251 & 0.253 \\ 
      LGBM & 0.246 & 0.247 & 0.247 & 0.247 & 0.248 & 0.249 & 0.250 & 0.249 & 0.250 & 0.251 \\ 
      CatBoost & 0.277 & 0.268 & 0.278 & 0.272 & 0.273 & 0.269 & 0.265 & 0.271 & 0.268 & 0.267 \\ 
      MLP & 0.263 & 0.264 & 0.265 & 0.266 & 0.269 & 0.271 & 0.271 & 0.272 & 0.275 & 0.276 \\ 
      TCN & 0.270 & 0.271 & 0.272 & 0.273 & 0.276 & 0.278 & 0.279 & 0.279 & 0.281 & 0.282 \\ 
      NBEATSx & 0.262 & 0.263 & 0.264 & 0.268 & 0.268 & 0.270 & 0.271 & 0.271 & 0.274 & 0.275 \\ 
      NHITS & 0.266 & 0.266 & 0.273 & 0.268 & 0.271 & 0.273 & 0.274 & 0.273 & 0.276 & 0.278 \\ 
    \bottomrule
    \end{tabular}
    \caption{M5 SMQL values for each method and retrain scenario combination.}
    \label{tab:m5_smql}
\end{table}

\begin{table}
    \centering
    \begin{tabular}{lrrrrrrrrrr}
    \toprule
    Method & 7 & 14 & 21 & 30 & 60 & 90 & 120 & 150 & 180 & 364 \\ 
    \midrule
      ETS & 0.105 & 0.105 & 0.105 & 0.104 & 0.103 & 0.102 & 0.102 & 0.101 & 0.103 & 0.102 \\ 
      LR & 0.110 & 0.110 & 0.110 & 0.111 & 0.111 & 0.111 & 0.112 & 0.112 & 0.112 & 0.113 \\ 
      XGBoost & 0.100 & 0.099 & 0.099 & 0.100 & 0.100 & 0.100 & 0.099 & 0.100 & 0.099 & 0.098 \\ 
      LGBM & 0.118 & 0.117 & 0.117 & 0.119 & 0.118 & 0.118 & 0.118 & 0.119 & 0.118 & 0.118 \\ 
      CatBoost & 0.156 & 0.157 & 0.154 & 0.157 & 0.157 & 0.157 & 0.157 & 0.156 & 0.156 & 0.155 \\ 
      MLP & 0.083 & 0.083 & 0.083 & 0.083 & 0.084 & 0.084 & 0.085 & 0.085 & 0.083 & 0.085 \\ 
      TCN & 0.083 & 0.083 & 0.083 & 0.082 & 0.082 & 0.082 & 0.082 & 0.082 & 0.080 & 0.080 \\ 
      NBEATSx & 0.084 & 0.084 & 0.084 & 0.088 & 0.085 & 0.085 & 0.085 & 0.086 & 0.088 & 0.105 \\ 
      NHITS & 0.087 & 0.085 & 0.101 & 0.086 & 0.087 & 0.088 & 0.088 & 0.091 & 0.088 & 0.089 \\   
    \bottomrule
    \end{tabular}
    \caption{M5 SMQC values for each method and retrain scenario combination.}
    \label{tab:m5_smqc}
\end{table}


\begin{table}
    \centering
    \begin{tabular}{lrrrrrrrrrr}
    \toprule
    Method & 1 & 2 & 3 & 4 & 6 & 8 & 10 & 13 & 26 & 52 \\ 
    \midrule
      ETS & 1.886 & 1.910 & 2.033 & 1.967 & 2.032 & 2.050 & 2.146 & 2.398 & 2.113 & 2.036 \\ 
      ARIMA & 1.935 & 1.944 & 1.969 & 1.996 & 1.897 & 1.789 & 2.010 & 1.898 & 1.821 & 1.830 \\ 
      LR & 6.886 & 6.873 & 6.925 & 6.886 & 6.892 & 6.888 & 6.870 & 6.947 & 7.037 & 7.077 \\ 
      RF & 1.871 & 1.873 & 1.872 & 1.868 & 1.867 & 1.866 & 1.861 & 1.863 & 1.852 & 1.840 \\ 
      XGBoost & 1.898 & 1.871 & 1.872 & 1.865 & 1.842 & 1.851 & 1.809 & 1.833 & 1.851 & 1.797 \\ 
      LGBM & 3.540 & 3.520 & 3.471 & 3.282 & 3.414 & 3.732 & 3.415 & 3.337 & 3.357 & 3.543 \\ 
      CatBoost & 5.631 & 5.570 & 5.694 & 5.025 & 4.846 & 5.359 & 4.418 & 4.876 & 5.212 & 5.343 \\ 
      MLP & 1.554 & 1.483 & 1.477 & 1.472 & 1.472 & 1.475 & 1.480 & 1.480 & 1.467 & 1.455 \\ 
      LSTM & 1.960 & 1.960 & 1.960 & 1.960 & 1.960 & 1.960 & 1.960 & 1.960 & 1.960 & 1.960 \\ 
      TCN & 1.960 & 1.960 & 1.960 & 1.960 & 1.960 & 1.960 & 1.960 & 1.960 & 1.960 & 1.960 \\ 
      NBEATSx & 1.704 & 1.696 & 1.661 & 1.561 & 1.519 & 1.525 & 1.544 & 1.462 & 1.455 & 1.456 \\ 
      NHITS & 1.705 & 1.680 & 1.629 & 1.556 & 1.547 & 1.460 & 1.528 & 1.453 & 1.455 & 1.451 \\ 
    \bottomrule
    \end{tabular}
    \caption{VN1 RMSSE values for each method and retrain scenario combination.}
    \label{tab:vn1_rmsse}
\end{table}

\begin{table}
    \centering
    \begin{tabular}{lrrrrrrrrrr}
    \toprule
    Method & 1 & 2 & 3 & 4 & 6 & 8 & 10 & 13 & 26 & 52 \\ 
    \midrule
      ETS & 0.156 & 0.114 & 0.093 & 0.070 & 0.054 & 0.039 & 0.032 & 0.034 & 0.014 & 0.000 \\ 
      ARIMA & 0.150 & 0.098 & 0.080 & 0.060 & 0.046 & 0.034 & 0.027 & 0.030 & 0.012 & 0.000 \\ 
      LR & 0.086 & 0.084 & 0.085 & 0.083 & 0.083 & 0.086 & 0.086 & 0.085 & 0.084 & 0.083 \\ 
      RF & 0.093 & 0.090 & 0.090 & 0.090 & 0.088 & 0.089 & 0.089 & 0.090 & 0.089 & 0.090 \\ 
      XGBoost & 0.146 & 0.129 & 0.125 & 0.120 & 0.118 & 0.115 & 0.114 & 0.119 & 0.116 & 0.113 \\ 
      LGBM & 0.121 & 0.110 & 0.111 & 0.106 & 0.104 & 0.105 & 0.105 & 0.104 & 0.104 & 0.105 \\ 
      CatBoost & 0.312 & 0.197 & 0.185 & 0.152 & 0.134 & 0.139 & 0.115 & 0.117 & 0.101 & 0.095 \\ 
      MLP & 0.457 & 0.468 & 0.467 & 0.475 & 0.442 & 0.483 & 0.437 & 0.492 & 0.468 & 0.451 \\ 
      LSTM & 0.557 & 0.301 & 0.202 & 0.126 & 0.101 & 0.044 & 0.055 & 0.048 & 0.032 & 0.013 \\ 
      TCN & 0.599 & 0.259 & 0.226 & 0.177 & 0.120 & 0.091 & 0.052 & 0.056 & 0.020 & 0.013 \\ 
      NBEATSx & 0.594 & 0.592 & 0.594 & 0.576 & 0.576 & 0.581 & 0.572 & 0.580 & 0.571 & 0.549 \\ 
      NHITS & 0.605 & 0.587 & 0.581 & 0.571 & 0.531 & 0.540 & 0.550 & 0.527 & 0.574 & 0.581 \\ 
    \bottomrule
    \end{tabular}
    \caption{VN1 sMAPC values for each method and retrain scenario combination.}
    \label{tab:vn1_smapc}
\end{table}

\begin{table}
    \centering
    \begin{tabular}{lrrrrrrrrrr}
    \toprule
    Method & 1 & 2 & 3 & 4 & 6 & 8 & 10 & 13 & 26 & 52 \\ 
    \midrule
      ETS & 1.003 & 1.032 & 1.077 & 1.084 & 1.115 & 1.133 & 1.159 & 1.249 & 1.196 & 1.259 \\ 
      ARIMA & 0.914 & 0.936 & 0.953 & 0.955 & 0.974 & 0.951 & 1.016 & 1.021 & 1.044 & 1.141 \\ 
      LR & 4.270 & 4.292 & 4.412 & 4.378 & 4.451 & 4.294 & 4.295 & 4.454 & 4.358 & 4.669 \\ 
      RF & 1.220 & 1.230 & 1.232 & 1.224 & 1.237 & 1.254 & 1.257 & 1.284 & 1.313 & 1.326 \\ 
      XGBoost & 1.503 & 1.496 & 1.493 & 1.505 & 1.483 & 1.488 & 1.466 & 1.451 & 1.515 & 1.479 \\ 
      LGBM & 2.318 & 2.292 & 2.230 & 2.157 & 2.230 & 2.317 & 2.262 & 2.362 & 2.220 & 2.268 \\ 
      CatBoost & 4.203 & 4.122 & 4.292 & 3.970 & 3.735 & 4.028 & 3.549 & 3.798 & 3.863 & 4.192 \\ 
      MLP & 0.806 & 0.800 & 0.796 & 0.795 & 0.803 & 0.805 & 0.807 & 0.819 & 0.824 & 0.847 \\ 
      LSTM & 0.892 & 0.895 & 0.899 & 0.903 & 0.910 & 0.917 & 0.924 & 0.935 & 0.957 & 1.003 \\ 
      TCN & 0.891 & 0.895 & 0.899 & 0.903 & 0.910 & 0.917 & 0.924 & 0.935 & 0.957 & 1.003 \\ 
      NBEATSx & 0.991 & 0.995 & 0.966 & 0.897 & 0.861 & 0.866 & 0.893 & 0.825 & 0.839 & 0.875 \\ 
      NHITS & 0.992 & 0.996 & 0.944 & 0.894 & 0.876 & 0.814 & 0.880 & 0.820 & 0.833 & 0.862 \\  
    \bottomrule
    \end{tabular}
    \caption{VN1 SMQL values for each method and retrain scenario combination.}
    \label{tab:vn1_smql}
\end{table}

\begin{table}
    \centering
    \begin{tabular}{lrrrrrrrrrr}
    \toprule
    Method & 1 & 2 & 3 & 4 & 6 & 8 & 10 & 13 & 26 & 52 \\ 
    \midrule
      ETS & 0.533 & 0.460 & 0.428 & 0.375 & 0.347 & 0.323 & 0.298 & 0.296 & 0.228 & 0.151 \\ 
      ARIMA & 0.463 & 0.396 & 0.345 & 0.304 & 0.291 & 0.260 & 0.255 & 0.235 & 0.182 & 0.132 \\ 
      LR & 1.559 & 1.551 & 1.545 & 1.536 & 1.549 & 1.525 & 1.544 & 1.632 & 1.674 & 1.894 \\ 
      RF & 0.729 & 0.732 & 0.735 & 0.737 & 0.735 & 0.738 & 0.743 & 0.734 & 0.731 & 0.734 \\ 
      XGBoost & 0.629 & 0.598 & 0.600 & 0.584 & 0.585 & 0.575 & 0.556 & 0.613 & 0.599 & 0.589 \\ 
      LGBM & 1.480 & 1.416 & 1.449 & 1.471 & 1.502 & 1.349 & 1.455 & 1.394 & 1.593 & 1.445 \\ 
      CatBoost & 2.265 & 1.859 & 1.765 & 1.582 & 1.546 & 1.577 & 1.329 & 1.576 & 1.302 & 1.243 \\ 
      MLP & 0.219 & 0.281 & 0.278 & 0.278 & 0.282 & 0.275 & 0.266 & 0.285 & 0.267 & 0.283 \\ 
      LSTM & 0.242 & 0.238 & 0.237 & 0.235 & 0.233 & 0.229 & 0.227 & 0.224 & 0.214 & 0.187 \\ 
      TCN & 0.241 & 0.238 & 0.237 & 0.235 & 0.233 & 0.229 & 0.227 & 0.224 & 0.214 & 0.187 \\ 
      NBEATSx & 0.765 & 0.763 & 0.708 & 0.567 & 0.477 & 0.442 & 0.520 & 0.348 & 0.362 & 0.369 \\ 
      NHITS & 0.755 & 0.779 & 0.660 & 0.555 & 0.506 & 0.333 & 0.507 & 0.348 & 0.343 & 0.365 \\  
    \bottomrule
    \end{tabular}
    \caption{VN1 SMQC values for each method and retrain scenario combination.}
    \label{tab:vn1_smqc}
\end{table}


\begin{figure}
    \centering
    \includegraphics[scale=0.65]{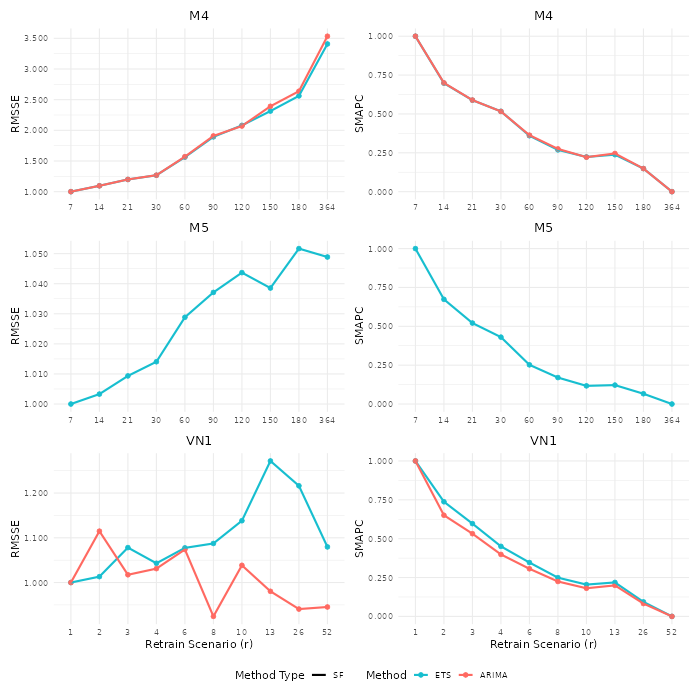}
    \caption{RMSSE and sMAPC results for each local method and retrain 
    scenario combination in relative terms with respect to the baseline 
    scenario, r = 7 for the M4 and M5 datasets, and r = 1 for the VN1 dataset.}
    \label{fig:local_rmsse_smapc}
\end{figure}

\begin{figure}
    \centering
    \includegraphics[scale=0.65]{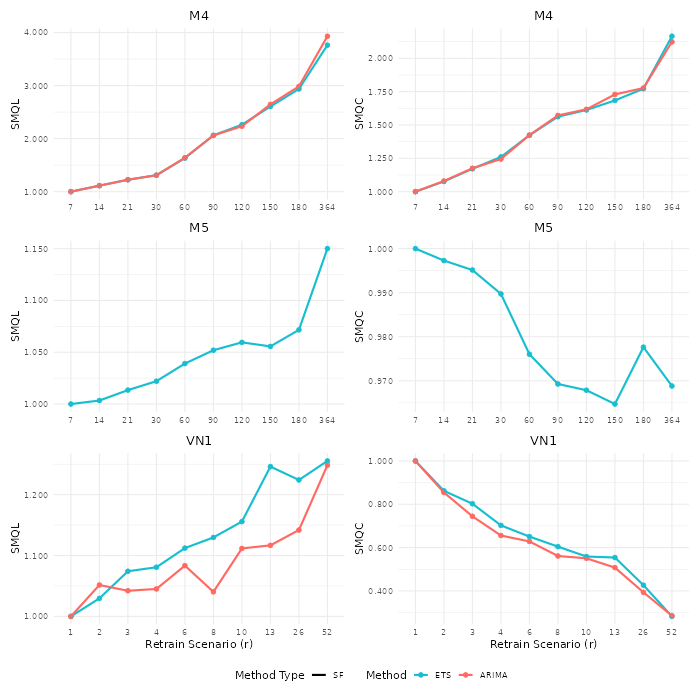}
    \caption{SMQL and SMQC results for each local method and retrain 
    scenario combination in relative terms with respect to the baseline 
    scenario, r = 7 for the M4 and M5 datasets, and r = 1 for the VN1 dataset.}
    \label{fig:local_smql_smqc}
\end{figure}


\begin{figure}
    \centering
    \includegraphics[scale=0.4]{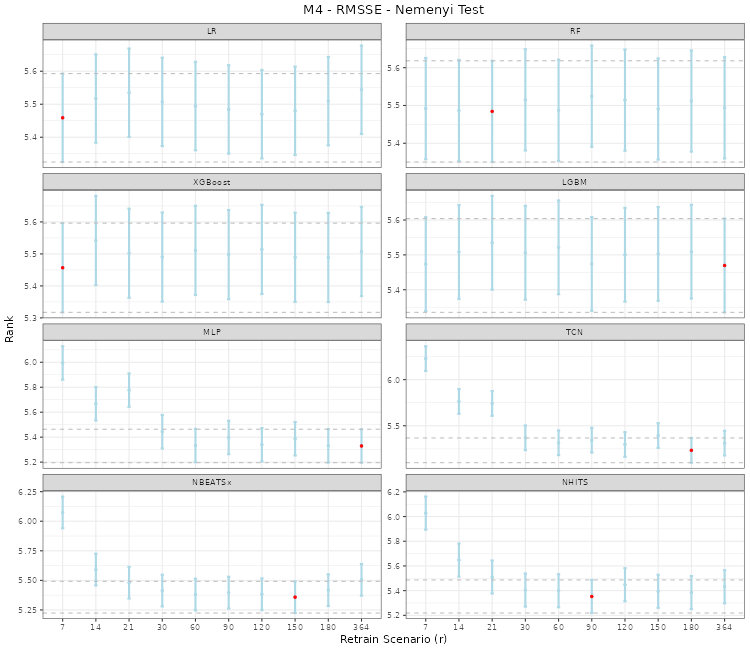}
    \caption{M4 Friedman-Nemenyi test results based on RMSSE.}
    \label{fig:test_mldl_m4_rmsse}
\end{figure}

\begin{figure}
    \centering
    \includegraphics[scale=0.4]{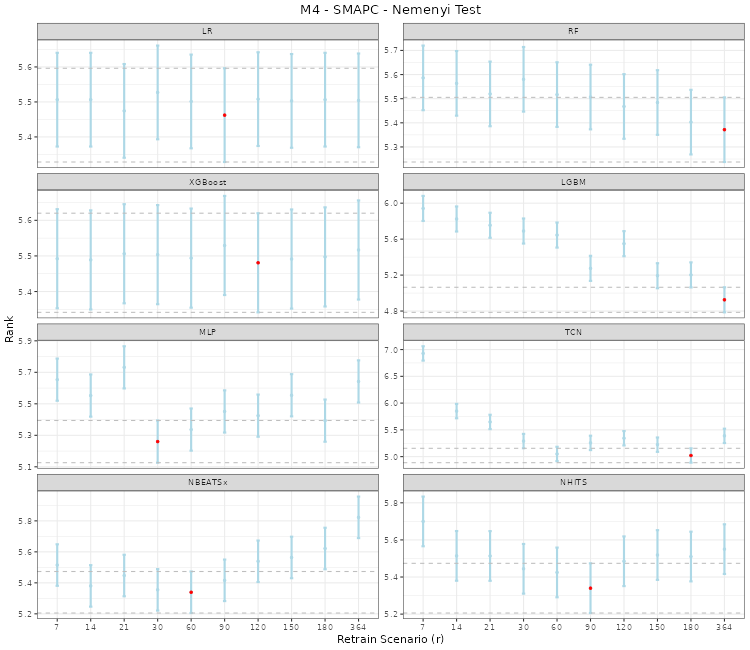}
    \caption{M4 Friedman-Nemenyi test results based on sMAPC.}
    \label{fig:test_mldl_m4_smapc}
\end{figure}

\begin{figure}
    \centering
    \includegraphics[scale=0.4]{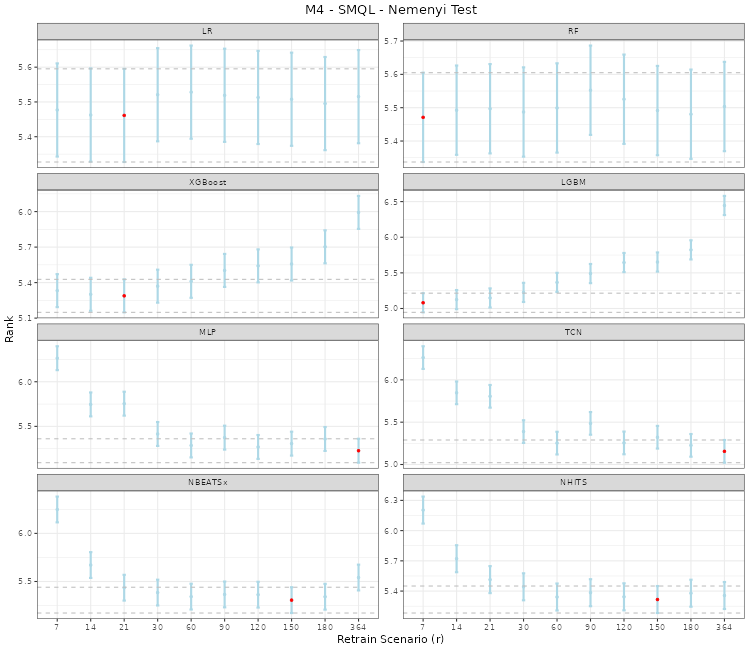}
    \caption{M4 Friedman-Nemenyi test results based on SMQL.}
    \label{fig:test_mldl_m4_smql}
\end{figure}

\begin{figure}
    \centering
    \includegraphics[scale=0.4]{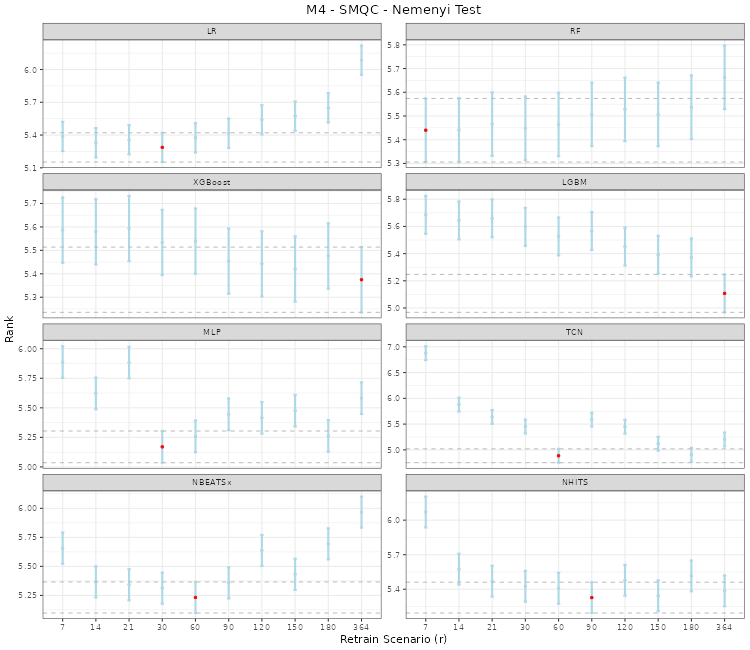}
    \caption{M4 Friedman-Nemenyi test results based on SMQC.}
    \label{fig:test_mldl_m4_smqc}
\end{figure}


\begin{figure}
    \centering
    \includegraphics[scale=0.4]{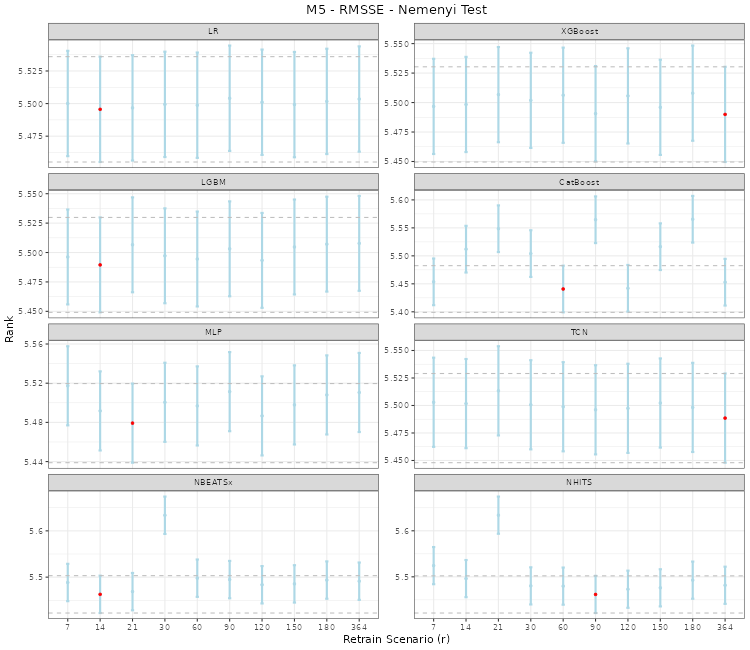}
    \caption{M5 Friedman-Nemenyi test results based on RMSSE.}
    \label{fig:test_mldl_m5_rmsse}
\end{figure}

\begin{figure}
    \centering
    \includegraphics[scale=0.4]{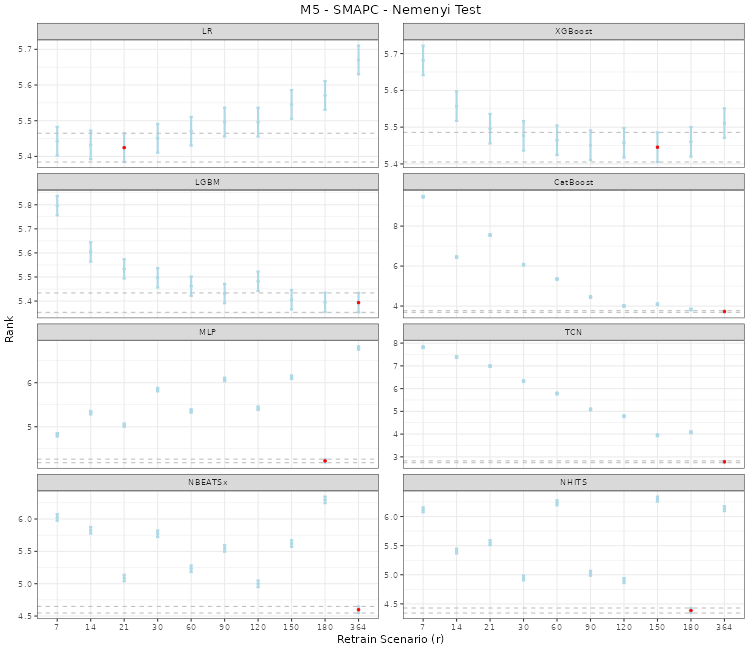}
    \caption{M5 Friedman-Nemenyi test results based on sMAPC.}
    \label{fig:test_mldl_m5_smapc}
\end{figure}

\begin{figure}
    \centering
    \includegraphics[scale=0.4]{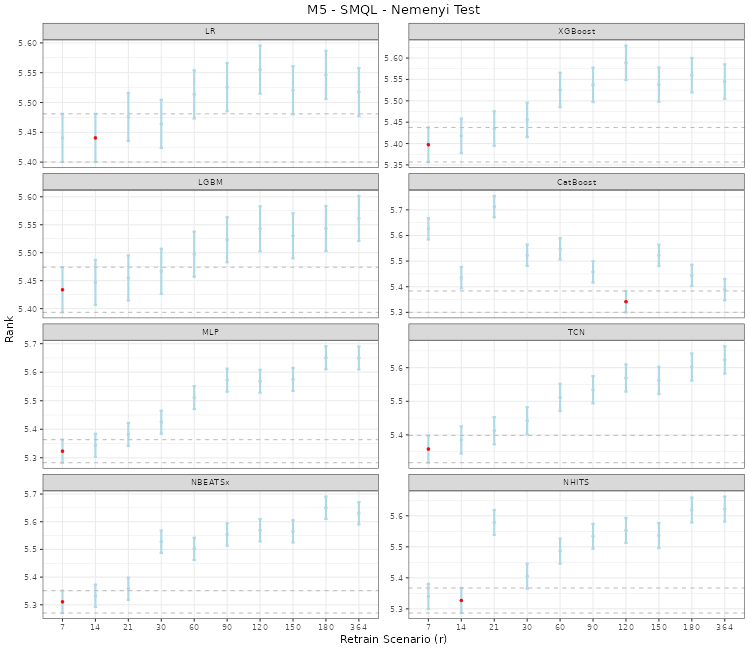}
    \caption{M5 Friedman-Nemenyi test results based on SMQL.}
    \label{fig:test_mldl_m5_smql}
\end{figure}

\begin{figure}
    \centering
    \includegraphics[scale=0.4]{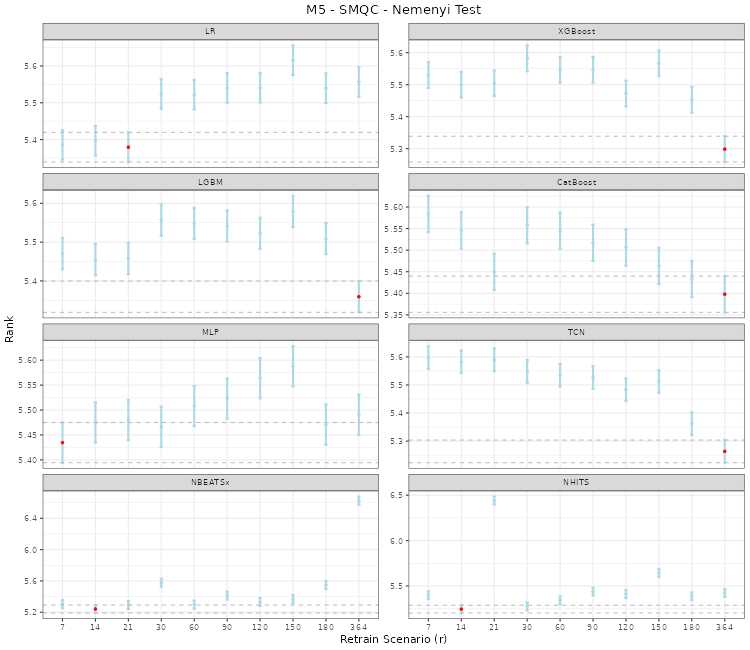}
    \caption{M5 Friedman-Nemenyi test results based on SMQC.}
    \label{fig:test_mldl_m5_smqc}
\end{figure}


\begin{figure}
    \centering
    \includegraphics[scale=0.4]{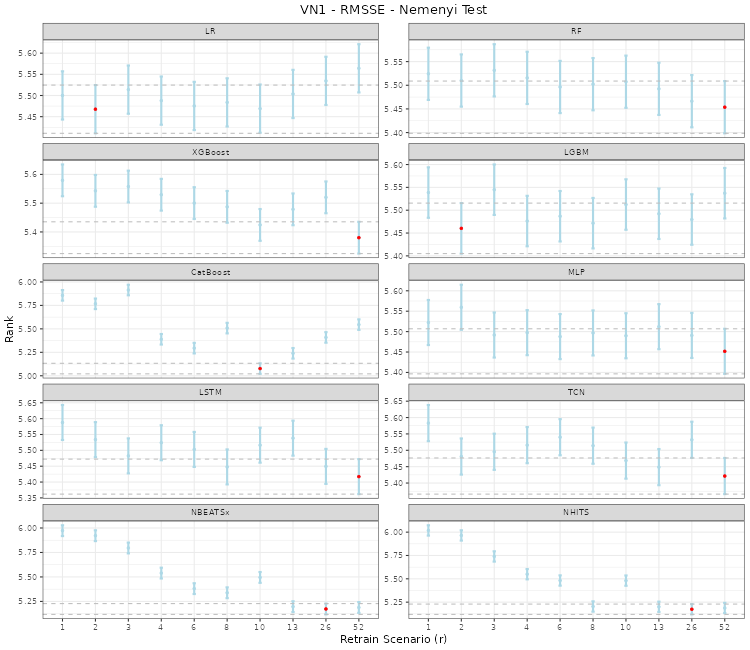}
    \caption{VN1 Friedman-Nemenyi test results based on RMSSE.}
    \label{fig:test_mldl_vn1_rmsse}
\end{figure}

\begin{figure}
    \centering
    \includegraphics[scale=0.4]{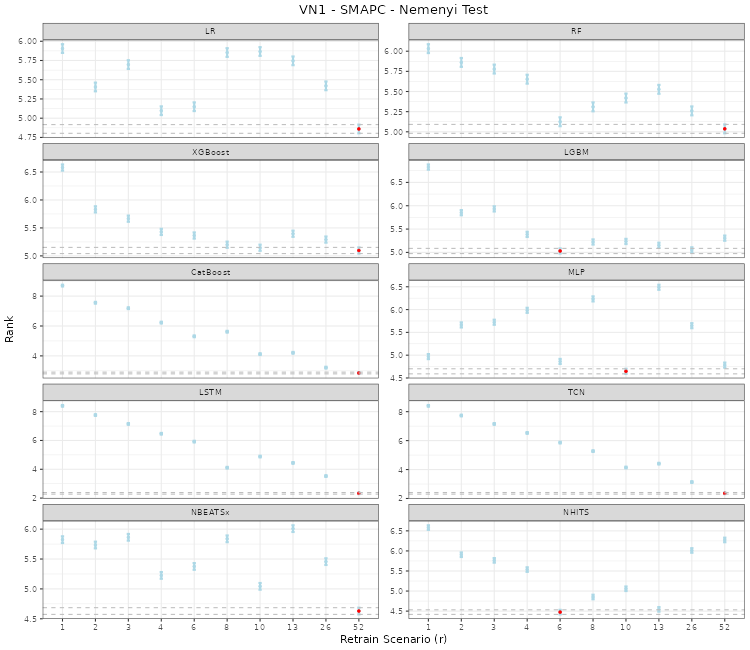}
    \caption{VN1 Friedman-Nemenyi test results based on sMAPC.}
    \label{fig:test_mldl_vn1_smapc}
\end{figure}

\begin{figure}
    \centering
    \includegraphics[scale=0.4]{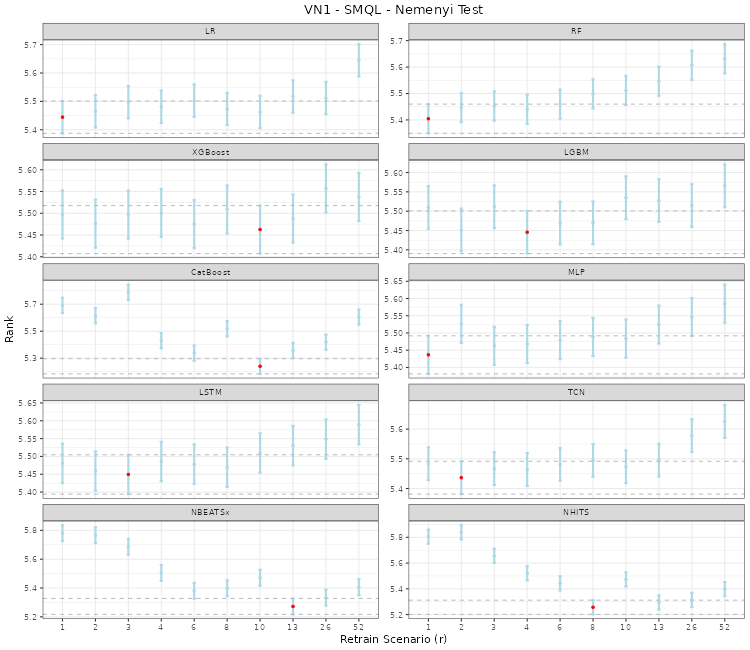}
    \caption{VN1 Friedman-Nemenyi test results based on SMQL.}
    \label{fig:test_mldl_vn1_smql}
\end{figure}

\begin{figure}
    \centering
    \includegraphics[scale=0.4]{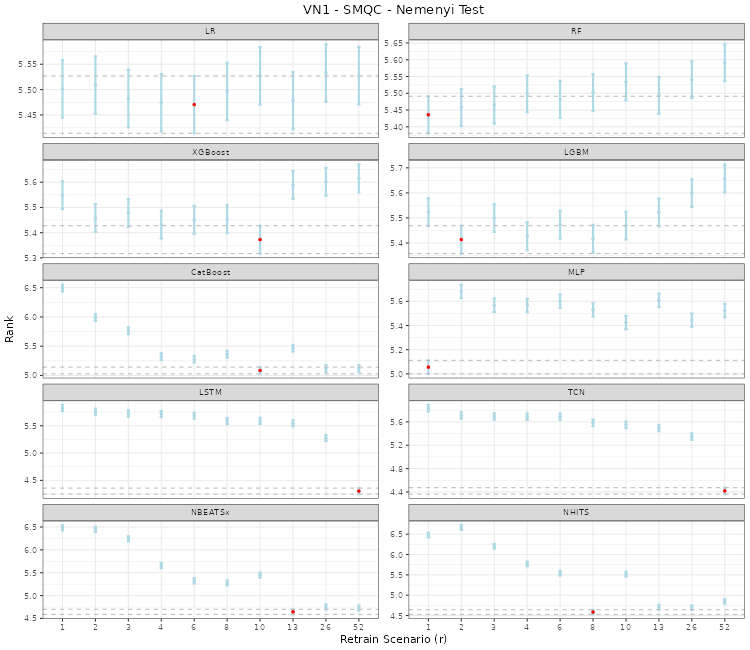}
    \caption{VN1 Friedman-Nemenyi test results based on SMQC.}
    \label{fig:test_mldl_vn1_smqc}
\end{figure}


\begin{figure}
    \centering
    \includegraphics[scale=0.6]{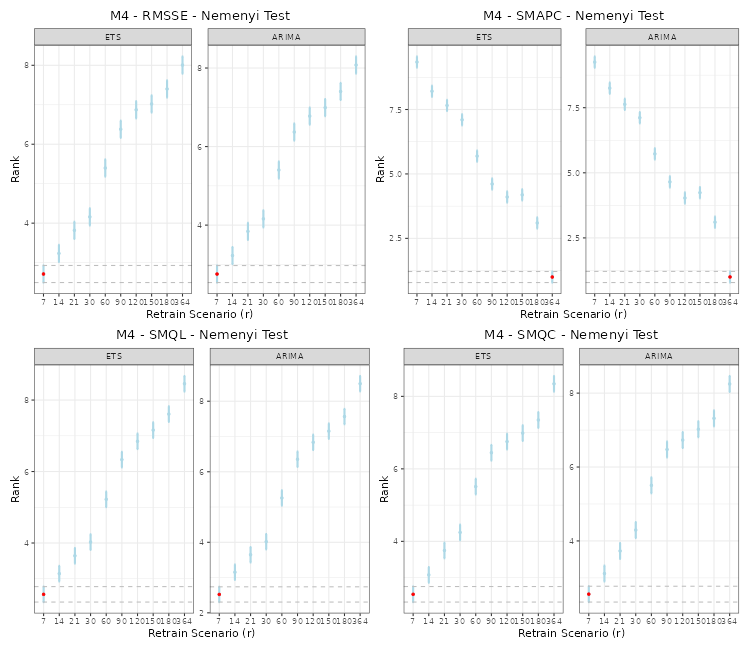}
    \caption{M4 Friedman-Nemenyi test results for local models.}
    \label{fig:test_sf_m4}
\end{figure}

\begin{figure}
    \centering
    \includegraphics[scale=0.6]{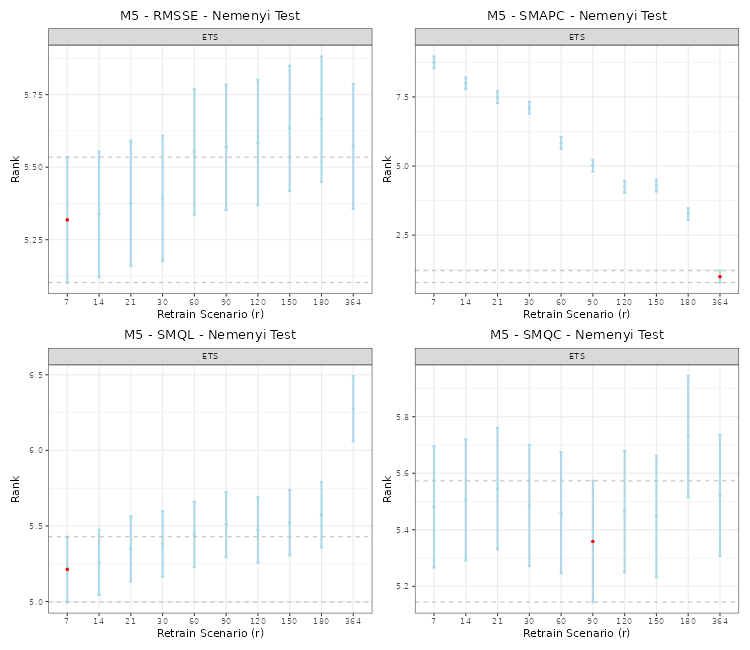}
    \caption{M5 Friedman-Nemenyi test results for local models.}
    \label{fig:test_sf_m5}
\end{figure}

\begin{figure}
    \centering
    \includegraphics[scale=0.6]{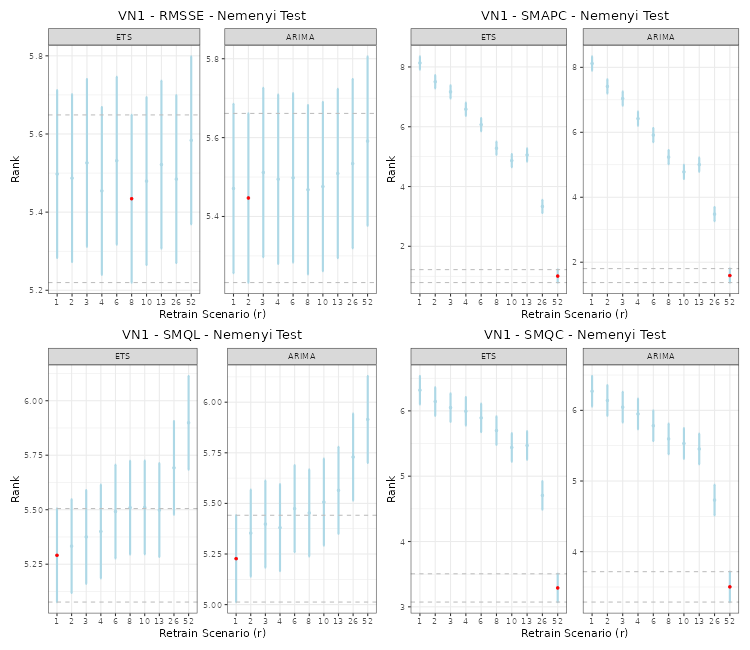}
    \caption{VN1 Friedman-Nemenyi test results for local models.}
    \label{fig:test_sf_vn1}
\end{figure}

\end{document}